\documentclass[twocolumn]{revtex4-1}
\usepackage{upgreek}
\usepackage{tikz}
\usepackage{dcolumn}
\usepackage{color,tcolorbox}
\usepackage{amsmath,mathrsfs,amssymb,esint}
\usepackage{dcolumn}
\usepackage{bm}
\usepackage{blindtext}
\usepackage{natbib}
\usepackage{graphicx}
\usepackage{rotating}
\usepackage{titlesec}
\usepackage[pdfborder={0 0 0},colorlinks=true,linkcolor=blue,urlcolor=blue,citecolor=blue]{hyperref}
\begin{document}
\title{Dynamical Equilibration Across a Quenched Phase Transition \\ in a Trapped Quantum Gas}

\author{I.-K. Liu$^{1,2}$, S. Donadello$^{3}$, G. Lamporesi$^{3}$, G. Ferrari$^{3}$, S.-C. Gou$^{2}$, F. Dalfovo$^{3}$, N.P. Proukakis$^{1*}$}

\affiliation{$^1$ Joint Quantum Centre (JQC) Durham--Newcastle, School of Mathematics, Statistics and Physics, Newcastle University,Newcastle upon Tyne, NE1 7RU, United Kingdom}
\affiliation{$^2$ Department of Physics ​and Graduate Institute of Photonic​s​, National Changhua University of Education, Changhua 50058, Taiwan}
\affiliation{$^3$ INO-CNR BEC Center and Dipartimento di Fisica, Universit\`a di Trento, 38123 Trento, Italy}

\begin{abstract}
The formation of an equilibrium quantum state from an uncorrelated thermal one through the dynamical crossing of a phase transition is a central question of non-equilibrium many-body physics. During such crossing, the system breaks its symmetry by establishing numerous uncorrelated regions separated by spontaneously-generated defects, 
whose emergence obeys a universal scaling law with the quench duration. Much less is known about the ensuing re-equilibrating or ``coarse-graining'' stage, which is governed by the evolution and interactions of such defects under system-specific and external constraints. 
In this work we perform a detailed numerical characterization of the entire non-equilibrium process, addressing subtle issues in condensate growth dynamics and demonstrating the quench-induced decoupling of number and coherence growth during the re-equilibration process. Our unique visualizations not only reproduce experimental measurements in the relevant regimes, but also provide valuable information in currently experimentally-inaccessible regimes.

\end{abstract}

\maketitle

The quenched crossing of a continuous second-order phase transition has been investigated both theoretically and experimentally in many physical systems. The prevailing scenario to date, known as the Kibble-Zurek mechanism \cite{Kibble76,Zurek85} describes the crossing of a phase transition under the assumption that, during the quench, the system dynamics is frozen. This mechanism was first proposed in cosmology by Tom Kibble \cite{Kibble76} to describe structure formation in the early Universe, but has been more widely adopted in the condensed matter context, thanks primarily to significant work by Wojciech Zurek \cite{Zurek85} and co-workers (see \cite{delCampo13} for a recent review).
Although it can be equivalently presented in terms of different physical observables, a common formulation relates the number of defects generated between regions of different (approximately constant) phase to the rate of external quenching.
In condensed matter, such a mechanism has been experimentally studied in superfluid helium \cite{Bauerle96,Ruutu96}, superconducting Josephson junctions \cite{Carmi00,Monaco06}, liquid crystals \cite{Chuang91,Fowler17}, multiferroic crystals \cite{Chae12,Lin14}, ions \cite{Ulm13,Pyka13,Ejtemaee2013} and ultracold atoms \cite{Weiler08,Chen11,Lamporesi13,Corman14,Navon15,Chomaz15,Donadello16,Aidelsburger17}. In the context of cold atoms specifically, which are ideal systems for generating and studying controlled non-equilibrium processes, recent experiments have already provided strong evidence for the Kibble-Zurek scaling through measurements of the number of spontaneously-generated defects in 3D harmonic traps \cite{Weiler08,Lamporesi13,Donadello16} or, equivalently, winding numbers in ring traps \cite{Corman14,Aidelsburger17}, with correlation function measurements in a 3D box-like potential used to extract critical exponents \cite{Navon15}, building on an extensive body of literature for condensate growth dynamics \cite{Kagan92,Stoof97,Miesner98,Gardiner98,Stamper-Kurn98,Bijlsma00,Stoof01,Davis02,Kohl02,Shvarchuck02,Hugbart07,Ritter07} -- see \cite{Davis16,Beu17} for recent reviews. Related quenched studies include soliton generation \cite{Zurek09,Damski10,Witkowska11}, ring-trap geometries \cite{Das12}, spinor \cite{Sadler06,Swislocki13,Saito13}, multi-component \cite{Sabbatini11,Liu16} and Josephson-coupled \cite{Su13} gases, and optical lattices  \cite{Chen11,Dziarmaga12,Braun15}.

A central emerging question is the extent to which the defect formation  process described by the Kibble-Zurek mechanism can be decoupled from the dissipative evolution expected to occur when defects co-exist within a finite region, which is crucial to the modelling of any experiment. Previous work on purely 1D evaporatively-cooled gases has indicated the spontaneous emergence of dark solitons which subsequently decay in time, with the coherence length of the final post-quench thermal state being set by the average distance between defects \cite{Witkowska11}. Decoupling these two dynamical features is not a trivial task, as one does not \textit{a priori} know exactly when to count the defects, and, in fact, at the time at which one is supposed to count the defects, the system is so disordered, that it is not even clear whether any counting would be possible, even at a post-processing stage. Nonetheless, numerous experiments with ultracold trapped atoms have provided convincing evidence of the emergence of the predicted power in the scaling of defect number with respect to quench rate, raising the question of why this scaling should survive over such prolonged dynamical periods. The Cambridge group \cite{Navon15} insightfully avoided such issues by looking instead at correlation functions, which enabled them to extract critical exponents, without the need for direct defect counting.

\begin{figure*}[t]
\includegraphics[width=1\linewidth]{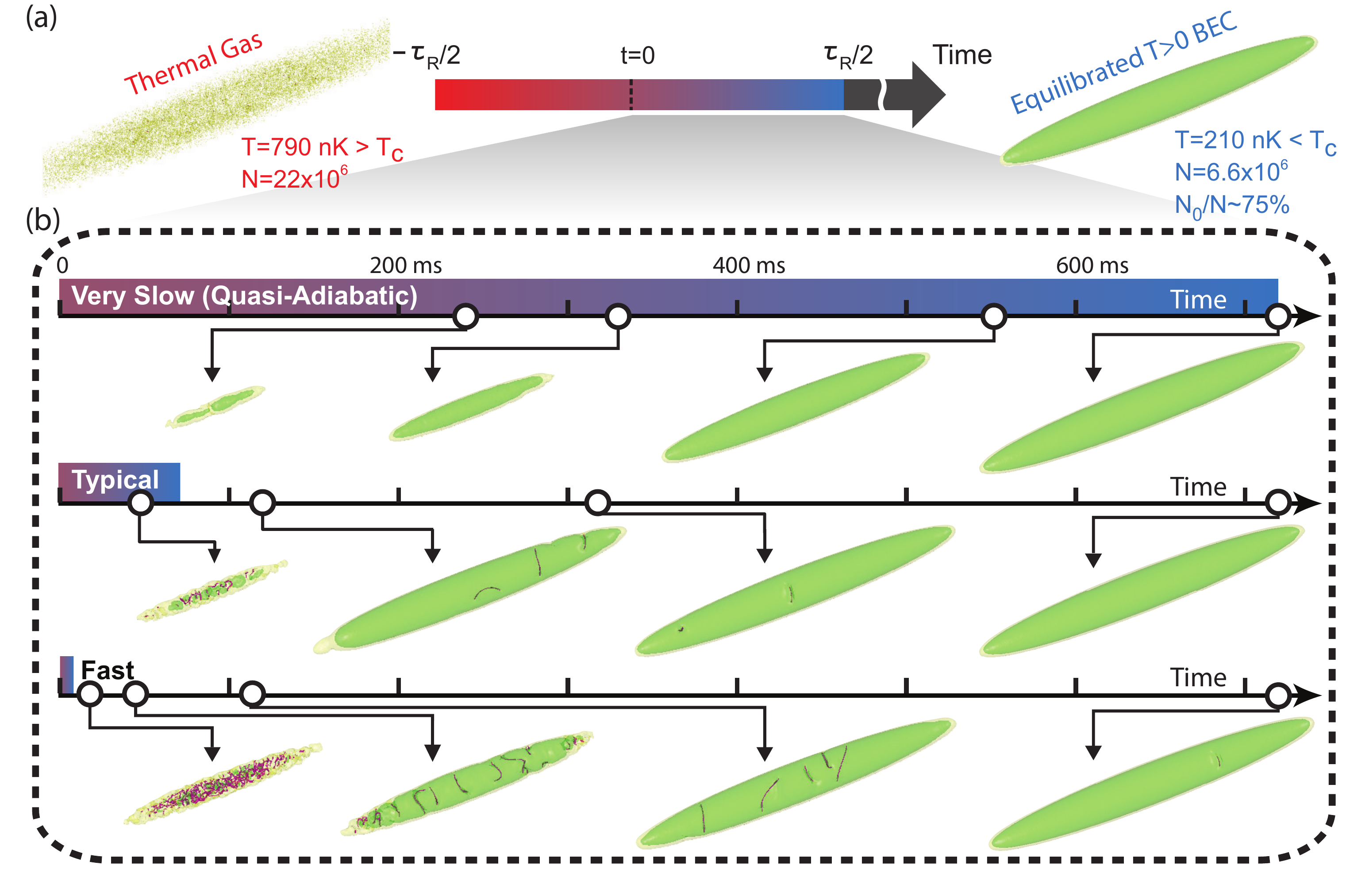}
\centering
\caption{ Simulation of quench-induced dynamical equilibration.
\textbf{a} Quench protocol: Starting from a purely thermal state, with a given atom number, we linearly quench temperature to lower values and chemical potential to positive values over a ramp duration $\tau_R$, to mimic experimental conditions.
\textbf{b} Dynamical response of an equilibrium thermal gas subjected to different cooling quench rates ($\tau_R = 1440, \, 144, \, 18$ ms, from top to bottom), demonstrating the equilibration route towards a finite temperature phase-coherent BEC. 
The characteristic regions depicted here refer to density isosurfaces of the highest-populated mode, chosen such that $n(t)/n(t \rightarrow \infty)$ = 0.1\% (yellow), or 3\% (green), where $n(t \rightarrow \infty)$  describes the final equilibrated condensate density for $N=6.6 \times 10^6$ atoms. 
Different rows correspond to different durations of the constant applied external cooling ramps, from the very slow, quasi-adiabatic (top) to the very fast, nearly-instantaneous ones (bottom), with the intermediate case representing typical quenches used in experimental studies of phase transitions. 
For rather slow ramps, most condensate formation happens during the external cooling. For shorter quench duration, the condensate appears around the end of the ramp, and there is a small number of spontaneously-generated defects.
In a fast quench, most condensate growth dynamics occurs after the end of the ramp, with the system at the end of the ramp being in a highly non-equilibrium state exhibiting large occupation of a handful of modes, and consisting of a dense random defect tangle. Such a tangle unravels in time into a phase-fluctuating condensate, or quasi-condensate, with numerous well-formed defects, whose presence perturbs the phase and opposes the formation of long-range coherence; the interacting defects gradually evolve into few long-living defects characteristic of the system geometry (quantized vortices here) which are experimentally observable after expansion, with the system asymptotically expelling all defects and acquiring phase coherence across its length.}
\label{fig_1}
\end{figure*}

In this work, we offer a unified analysis of quenched condensate growth dynamics in a finite elongated 3D inhomogeneous system,   
incorporating into a single study the dynamical evolution, for different induced quench rates, from an equilibrium thermal state above the Bose-Einstein condensation transition temperature to a near-equilibrated, low-temperature phase-coherent Bose-Einstein condensate (BEC). 
By demonstrating the natural emergence of symmetry-breaking in our simulations, we perform a detailed parallel analysis of both the spontaneous emergence and complex nonlinear dynamics of defects, and the related evolution of coherence.
Addressing subtle open questions in the condensate formation process, we show that coherence and number growth dynamics are in general decoupled, due to competing growth mechanisms following a quench, except for cases of adiabatically-slow growth which exhibit broadly-similar timescales. The unique insight provided by our numerical visualizations, which extend beyond experimentally-accessible time intervals, provides a natural framework for addressing the still unresolved interplay between Kibble-Zurek defect generation and coarse-graining dynamics. In this regard, our simulations demonstrate that the anticipated power-law defect scaling with quench rate (characteristic of Kibble-Zurek) is not significantly affected over a prolonged evolution period, despite the fact that both the number and type of defects changes considerably during this period. Specifically, the initial randomly-oriented and -shaped spontaneously-generated defects gradually relax to the canonical excitations in the given system geometry (here solitonic vortices), with their number exhibiting a rapid decrease at very early stages, followed by a more gradual decay process as the size of the condensate grows further. These findings are consistent with late-time experimental measurements performed within our group \cite{Lamporesi13,Serafini15,Donadello16}, which further validate our analysis.

In order for our predictions to be directly comparable to experimental measurements in the appropriate regimes, we perform full 3D numerical simulations of the stochastic (projected) Gross-Pitaevskii equation \cite{Blakie08,Proukakis08,Proukakis13,Weiler08,Su13,Liu16,Stoof01,Cockburn10} in the elongated configuration of our recent experiments \cite{Lamporesi13,Serafini15,Donadello16}. Such a regime is particularly interesting as it facilitates unique access to the role of coherence in a finite-size, inhomogeneous and anisotropic system, for which the phase transition process becomes position-dependent, with the condensate first emerging in the central trap regions, where the phase-space density is maximised.
Examples of simulations of the BEC growth for different quench rates are given in Fig.~\ref{fig_1}, clearly showing the dependence of defect generation and subsequent dynamical equilibration on quench rate.  

\section{Results}

\subsection{Quench Protocol}

Our experiment is conducted in a cigar-shaped trap with $\omega_x/2 \pi=13$~Hz and $\omega_\perp/2 \pi=131.4$~Hz, where 
$^{23}$Na atoms in the $|F,m_F\rangle =|1,-1 \rangle$ state are evaporatively cooled across the BEC phase transition at different rates. To faithfully mimic both changing temperature and total atom number observed in the experiments, our simulations are based on linear quenches (Fig.~\ref{fig_1}a) over a finite quench duration $\tau_R$ , both in temperature ($T = 790$~nK $\rightarrow 210$~nK) and chemical potential ($\mu = -22\, \hbar \omega_\perp \rightarrow\mu = + 22\, \hbar \omega_\perp$).
Consistently, the atom number goes down from  $N=22 \times 10^6$ to $N=6.6 \times 10^6$. Our parameters have been chosen such that $t=0$ corresponds to the time when the system crosses the ideal gas critical temperature ($\mu=0$).

For each quench rate, the results are analysed over 3 to 7 individual realisations, which are sufficient for understanding the underlying physics. Our analysis is based on the characterization of the emerging condensate, defined in our simulations as the mode with the largest eigenvalue of the one-body density matrix, based on the usual Penrose-Onsager definition  \cite{Penrose56,Blakie08}.  We also note that at early evolution times there are a number of approximately-equally largely populated modes, before one becomes randomly dynamically favoured by the system. Details of our experimental configuration have been reported elsewhere \cite{Lamporesi13,Donadello16}, while the stochastic numerical method and data analysis scheme used to model such non-equilibrium dynamics are summarised in {\em Methods}. 

\begin{figure*}[ht!]
\includegraphics[width=1\linewidth]{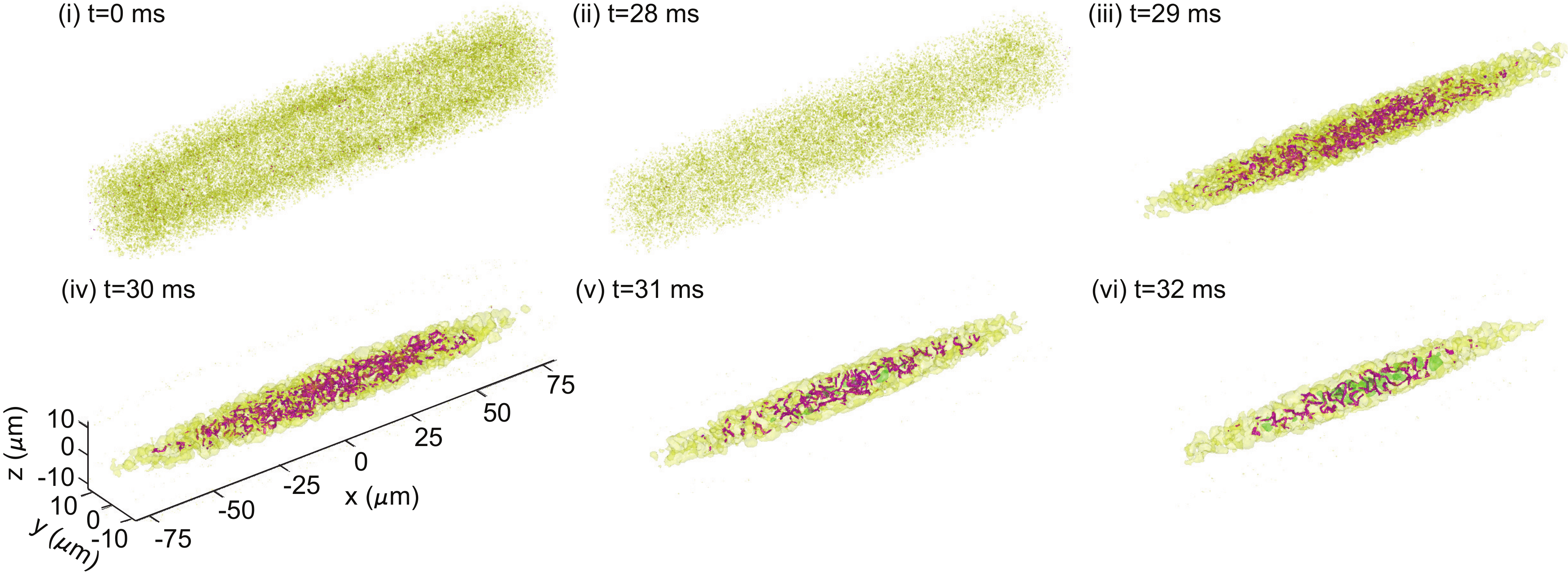}
\centering
\caption{ Dynamical crossing of the BEC phase transition. We show here a 3D visualization of characteristic density profiles of the highest-occupied (Penrose-Onsager) mode during the condensation process, for the case of quench time $\tau_R=84$~ms. Above the transition temperature, we see low-amplitude random density fluctuations (yellow) within our entire simulation grid, with the density condensing in real space to reflect the underlying elongated trap geometry. During such condensation  process, the system breaks its symmetry spontaneously, leading to the appearance of defects (purple) within a growing high-density region (green), signalling the emergence of a highly-fluctuating condensate. Such defects undergo their own evolution dynamics, violently perturbed by interactions with other defects (Supplementary Fig.~3 and \href{https://youtu.be/w-O2SPiw3nE}{Movie 3}).}
\label{fig_2}
\end{figure*}

Fig.~\ref{fig_1}b shows that for a very slow quench (top), the system grows gradually to its final equilibrium state. In doing so it remains close to the corresponding equilibrium phase-coherent BEC during its evolution, except in a narrow time window within the region of critical fluctuations. Such slow evolution corresponds to the evaporatively-cooled condensate growth scenario, in which both condensate atom number and correlation function grow gradually, with a small phase-coherent condensate present shortly after the phase transition. In such cases, the system equilibration happens in real time, on the ramp, i.e., during evaporative cooling. As the ramp speed is increased (middle and bottom images), we observe more clearly a rather violent falling out of equilibrium of the dynamical system when crossing the phase transition, resulting in the spontaneous production of multiple defects: the underlying physical picture here is that coherence only forms in local patches of constant -- but random -- phase, and the defects separate such regions of different phases: Faster quenches (bottom) lead to a faster overall initial growth in the atom number of the quantum-degenerate state, and a larger number of spontaneously-generated defects than slower quenches (middle), with most of the condensate (and also coherence) growth occuring after the removal of the external ramp.

As the system grows, such defects are stretched out while also undergoing complicated nonlinear dynamics in the inhomogeneous background, where they  interact with other defects. As a result of this, the number of defects decreases gradually (with the rate of decrease depending critically on quench rate), allowing phase coherence to spread to the entire system.
For the set of experimental parameters considered here, the final equilibrated state -- obtained asymptotically for the case of near-instantaneous ramps -- consists of a fully phase-coherent defect-free finite-temperature BEC with condensate fraction $N_0/N \approx 0.75$. 
(Characteristic single-realisation growth sequences are shown in Supplementary Fig.~\ref{fig_1}-\ref{fig_2} and \href{https://youtu.be/3q7-CvuBylg}{Movies 1}-\href{https://youtu.be/-Gymaiv9rC0}{2}).

In the following section we analyse in detail such distinct stages in the dynamical equilibration, offering further insight to this complex non-equilibrium process, while also demonstrating very good agreement between theory and experiment, for the observables and regimes where experimental measurements have been undertaken.

\subsection{Dynamical Crossing of the Phase Transition}

The transition region dynamics is naturally incorporated within our numerical approach. As the system is cooled at a finite rate, with its atom number decreasing (as in experiments), it cannot instantaneously acquire coherence across its entire spatial extent, and so a dynamical spontaneous symmetry-breaking occurs over a relatively short temporal range; this allows the system to become infilled by a densely-packed network of randomly-located defects shown in purple around $t\approx30$ ms. The emergence of such defects signals a stark deviation from the corresponding equilibrium system profiles. This process is demonstrated in Fig.~\ref{fig_2} for the specific example of a quench time $\tau_R = 84 $~ms. 

Consistent with experiments, our simulations clearly show the temporal narrowing of the density distribution during the phase transition, resulting in the gradual emergence of a higher-density condensate region (green region, $t \ge 31$~ms) which gradually grows towards the trap edges. The origin of this is associated with the lowering of the system temperature, the incorporated atom loss and the presence of the harmonic confinement, all of which lead to a centre-peaked position-dependent increase in local phase-space density. However, looking at the finer level during such evolution, we find the defects also interact, coalesce and decay. Such processes, which are {\em not} included in the Kibble-Zurek mechanism are nonetheless crucial to the growth of the phase-coherent regions: In the early stages, while numerous defects are present, the system can be classed as a quasi-condensate \cite{Kagan92,Petrov01} in the sense that different regions of coherent density exhibit no common phase between them, such that the observed coherence length is considerably smaller than the system size, consistent with the large population of a number of modes. 

Our advanced numerical visualisation, which encapsulates the phase transition physics, reveals the complexity of attempting to extract, whether numerically or in actual experiments, the early physics of defect formation. This is of particular relevance for addressing the interplay between the simplified Kibble-Zurek model and the ``coarse-graining'' dynamics governing dynamical defect evolution and decay. Our analysis shows that one cannot decouple the two processes -- at least not in the context of inhomogeneous systems.

Following the crossing of the phase transition and the spontaneous formation of defects, the emerging condensate enters into a rather distinct dynamical regime, in which the physics is dominated by a complicated nonlinear combination of (i) defect stretching due to the growing condensate size, (ii) defect propagation in an inhomogeneous environment, (iii) occasional but rather violent defect interactions, including vortex reconnections, bouncing and ``ejection'' from the condensate \cite{Serafini17}, and (iv) additional forced relaxation in cases of slow cooling. The combination of such processes leads to the gradual dynamical equilibration of the system, which evolves from a defect-filled condensate to a state at $T \ll T_C$ defined by the final quench parameters.

\begin{figure}[t]
\includegraphics[width=1\linewidth]{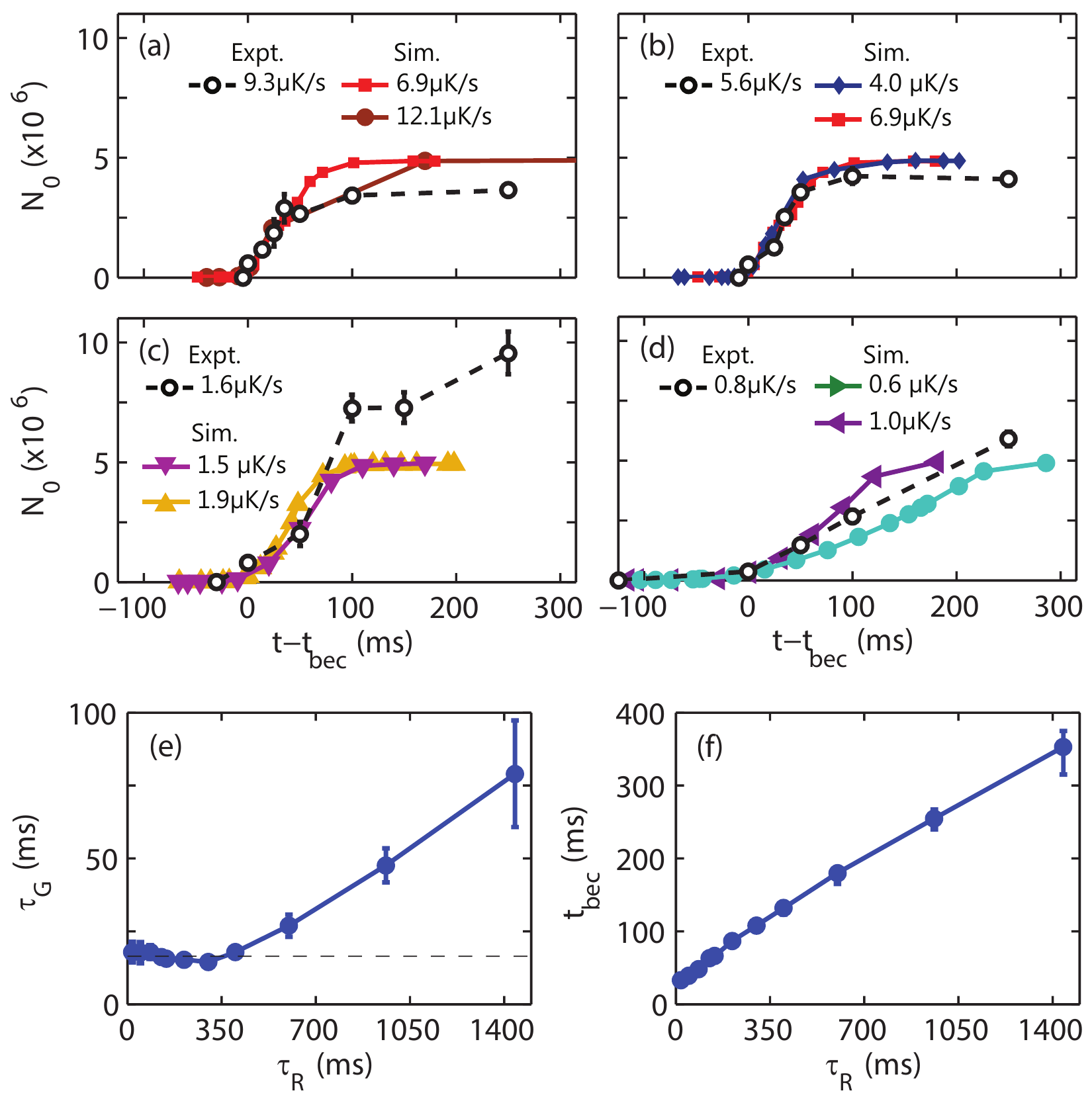}
\centering
\caption{ Condensate Growth Dynamics. \textbf{a-d} Comparison of dynamical simulations (filled symbols, solid lines) and experimental data (open black circles, dashed lines) for different cooling rates, labelled here by the rate of change of temperature with time ($dT/dt$). Each subplot shows a complete averaged experimental sequence of condensate atom number $N_0$  for a particular quench rate, with the depicted averaged numerical results corresponding to quench rates either side of the experimental values. 
In the simulations we assume a fixed initial ($T=790$nK $> T_C$, $N = 22 \times 10^6$ atoms) and final equilibrium ($T=210$nK $< T_C$, $N = 6.6 \times 10^6$ atoms) states. We also fit the numerical data with an S-shaped function (see Eq.~(\ref{eq:sshape}) in {\it Methods}) to extract the growth timescale $\tau_G$. 
\textbf{e} Dependence of $\tau_G$ on the ramp duration $\tau_R$ in the simulations, with  error bars representing the $95\%$ confidence bounds of the fits. 
\textbf{f}  Condensation onset time, $t_{\rm bec}$, as a function of $\tau_R$, demonstrating that slower quenches feature a delayed onset. Error bars here represent the difference in time for the condensate atom number to reach 3\% (lower bound) or 7\% of the total final atom number.  }
\label{fig_3}
\end{figure}

\subsection{Condensate Growth Dynamics}

Early condensate growth experiments \cite{Miesner98,Ritter07,Hugbart07} and their numerical modeling \cite{Gardiner98,Bijlsma00,Davis02} revealed a number evolution $N_0(t)$ well described by an S-shaped curve. However, slower quenches \cite{Kohl02} and elongated geometries \cite{Dettmer01} were observed to feature a pronounced region of critical fluctuations, leading to a ``time delay'', or ``onset time'' for condensate growth and a slower (near-linear) initial growth rate \cite{Kohl02}; although the presence of such features is well-documented and broadly attributed to the initial emergence of the quasicondensate, there has been little quantitative discussion of this issue, which becomes particularly relevant for dynamically-driven quenches.

To address this point, we firstly compare our numerical and experimental condensate growth curves for different characteristic temperature quench rates $dT/dt$, as shown in Fig.~\ref{fig_3}.  Consistent with earlier numerical studies of condensate growth (including Kibble-Zurek dynamics \cite{Weiler08,Liu16}), the constant $\gamma$ appearing in our theoretical model (see Eq.~(2) in {\it Methods}) is treated as a free parameter, choosing its value such that it reproduces the experimental growth rate for $dT/dt=5.6$~$\mu$K/s (Fig.~\ref{fig_3}b).   
In doing so, our numerical results are shown to accurately reproduce experimental growth curves for the entire range of quenches probed.

The precise determination of the phase-transition crossing and associated critical temperature is a rather challenging problem, both numerically and experimentally, particularly in the presence of inhomogeneous confinement.
Experiments typically identify the critical region as the time at which a clearly detectable condensate emerges. 
Following such a protocol, in our numerical simulations we identify a corresponding condensation ``onset time'', $t_{\rm bec}$ during the early condensate growth stages, as the time at which the condensate atom number, $N_0$, -- defined as the largest eigenvalue of the one body density matrix -- reaches $5\%$ of the total final atom number for our chosen final equilibrium parameters; such definition is broadly consistent with our experimental measurements.
Shifting our numerical growth curves by the ``onset'' time $t_{\rm bec}$ enables a direct comparison to experimental growth curves, with the good agreement shown in Fig.~\ref{fig_3}a-d.

All our numerical condensate growth curves are well fitted by an S-shaped curve with a {\em single} free parameter, corresponding to the quenched growth timescale, $\tau_G$ (Fits shown in Supplementary Fig.~\ref{fig_4}).
The dependence of $\tau_G$ and condensate onset time, $t_{\rm bec}$ on ramp duration is shown in Fig.~\ref{fig_3}e,f. For $\tau_R \lessapprox 300$~ms, we find a practically constant growth timescale $\tau_G=15.8 \pm 0.7$~ms, which is consistent with the finding of overlapping condensate growth curves once these are plotted against shifted time $(t-t_{\rm bec})$.
Note that the condensate onset time  $t_{\rm bec}$ is a linear-like  monotonically-increasing function of $\tau_R$.
This behaviour is qualitatively robust to changes in the exact definition of the condensate number/fraction chosen to mark such transition, as demonstrated by the small error bars in Fig.~\ref{fig_3}f.

\subsection{Defect Number Dynamics and Visualization}

\begin{figure}[t!]
\includegraphics[width=1\linewidth]{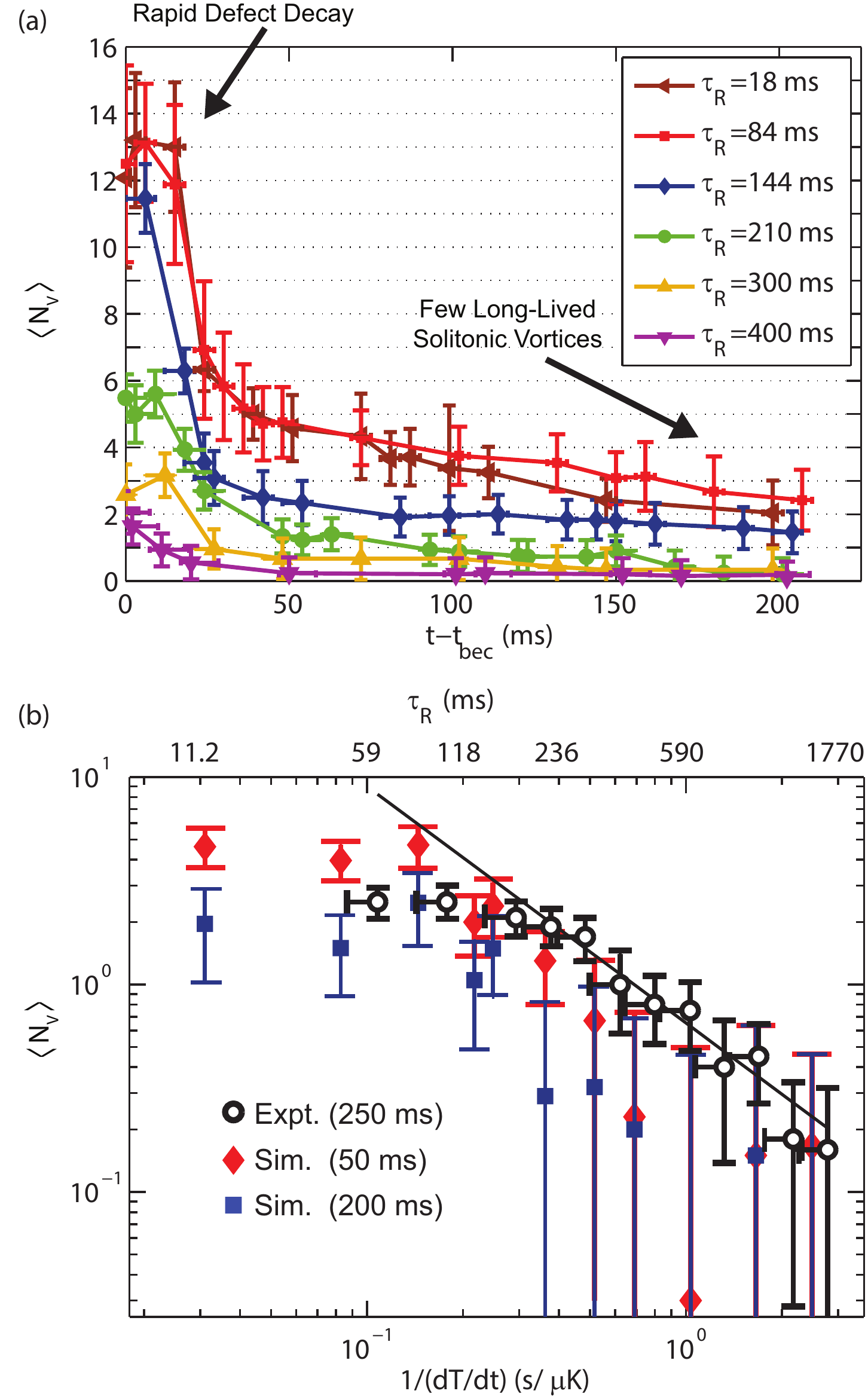}
\centering
\caption{  Mean Defect Number Dynamics and Dependence on Cooling Ramp.
\textbf{a} Evolution of mean defect (vortex) number, $N_v$, after the onset of condensation for different $\tau_R$, highlighting the initial rapid defect decay regime ($t-t_{\rm bec} \lessapprox 30$~ms) and the subsequent regime where few vortices prevail for an extended amount of time. 
\textbf{b} Dependence of mean defect number on inverse cooling rate $(dT/dt)^{-1}$ for two different simulated evolution times $t-t_{\rm bec}$ $\approx 50$ ms (red diamonds), and $\approx$ 200 ms (blue squares) compared to corresponding experimental  values after a free evolution time of 250 ms (black circles). The solid line is a linear fit to the experimental data in log-log scale in the slow quench region (last nine points). The emergence of a plateau-like region on the left (quenches with $\tau_R \lessapprox 144$ ms, or high $dT/dt$) is consistent with the saturation of the number of long-lived defects seen for long simulation times in (a).  Error bars in vortex numbers represent statistical uncertainties (see {\it Methods}); horizontal error bars in (a) arise from the uncertainty in the determination of $t_{\rm bec}$. 
}
\label{fig_4}
\end{figure}

\begin{figure*}[t]
\includegraphics[width=1\linewidth]{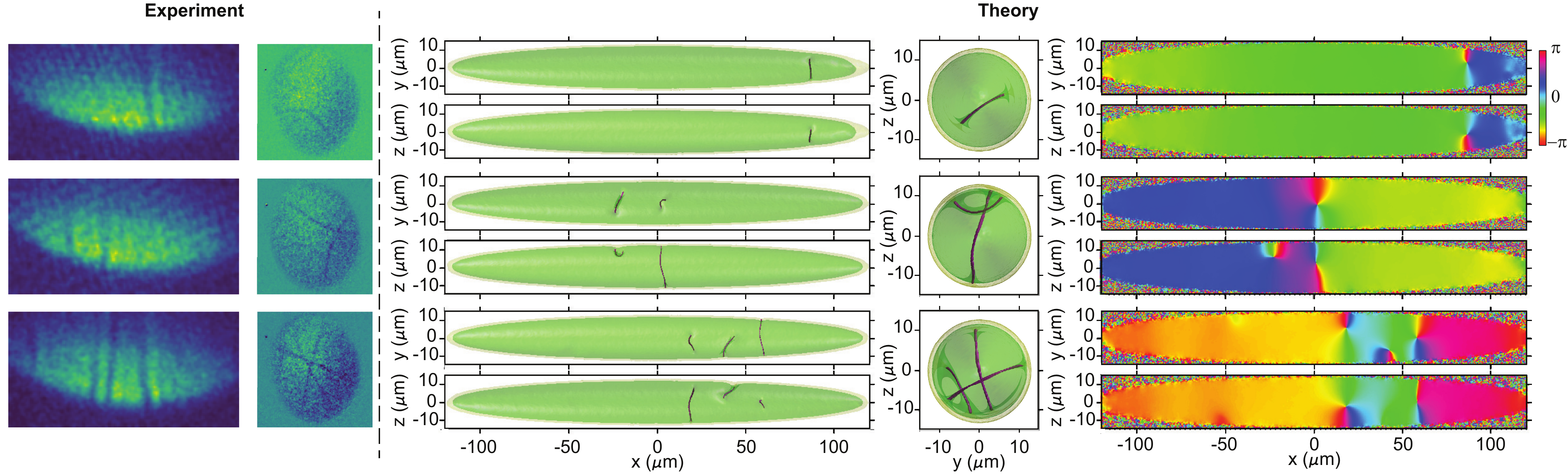}
\centering
\caption{ Defect visualization at late evolution times. Shown here are representative density side views containing respectively (top to bottom) 1, 2 and 3 solitonic vortices in the condensate.
Left Images: experimental densities following condensate expansion, integrated along a transverse (1st column) or longitudinal (2nd column) axis;
Right images: numerical simulations showing corresponding characteristic density and phase plots: simulations offer both a view of two different planar cuts (so at $z=0$, or $y=0$), and the corresponding planar phase profiles, which clearly highlight the solitonic-vortex nature of the defects at late evolution times. Simulation data correspond to $\tau_{R}=84$ ms and an evolution time $(t-t_{\rm bec})\approx 250$ ms.
}
\label{fig_5}
\end{figure*}

Next, we discuss the number (Fig.~\ref{fig_4})  and nature (Fig.~\ref{fig_5}) of the emerging defects restricting our analysis to times $t \ge t_{\rm bec}$, with a qualitative analysis at earlier times prohibited by the densely-packed defect configurations present at early post-quench evolution times.

{\em Defect Generation and Evolution:} 
The number of defects already present at $t_{\rm bec}$ varies significantly with ramp duration, being much higher for faster ramps, as such ramps create a more non-equilibrium initial configuration for the system.
Despite the difference in absolute numbers, all defect-number curves exhibit a similar dynamical behaviour, as evident in Fig.~\ref{fig_4}a. During the first $\approx 30$~ms, such curves exhibit a rather rapid initial decrease in their numbers associated with defect interactions. This is followed by a period of slower decrease with just a few defects present in the system. Such defects interact only occasionally, and mostly in pairs, because the typical distance between them during their motion is relatively-large. 
However, during such interactions they can violently emit a significant amount of energy, and occasionally/eventually get converted to a sound wave, or ejected by the evolving condensate near the edges of the system. 
Consistent with previously-reported experimental findings \cite{Serafini15,Serafini17}, for which the defect number could only be reliably measured after about a hundred ms of in-trap evolution, we find the decay in the number of defects to be significantly slower once only two defects are left in the system in any given individual run; in that case, their dynamics becomes largely decoupled, dominated by free propagation in the inhomogeneous condensate. This latter evolution stage is consistent with exponential decay, on a timescale of one hundred to few hundred ms.

 {\em Relation to Kibble-Zurek Mechanism:}
We next investigate (Fig. \ref{fig_4}b) the mean number of defects (vortices) as a function of ramp duration (top axis) or, equivalently, inverse temperature evolution (bottom axis). According to the idealised Kibble-Zurek scenario for defect generation, one would expect a power-law decay with a particular exponent.
However, such Kibble-Zurek predictions are specific for a relatively early evolution time within the critical region, and are based on a scenario of frozen dynamics  which does not account for the possibility of defect interactions, whereas experiments typically count defects after a significant in-trap evolution time and subsequent expansion imaging. Our simulations have already shown (Fig.~\ref{fig_4}a) that the defect number significantly changes during the in-trap evolution. Nonetheless, the power-law scaling of the defect decay (i.e.\ the slope of the decaying region in Fig.~\ref{fig_4}b) appears not to be very sensitive to the exact post-quench timing of the measurement, thus demonstrating the consistency of our numerically-identified defect dynamics with the experimental findings~\cite{Donadello16}. Specifically,  
a detailed analysis of defect number vs. quench rate for two different evolution times $(t-t_{\rm bec}) \sim 50$ ms (red diamonds) and 200 ms (blue squares), reveals a power-law decay consistent with both the predicted range and the experimental measurements (conducted at 250~ms).
Our simulations also recover the experimentally-observed plateau region for fast quenches \cite{Donadello16}, a behaviour which we find to set in already at early post-quench times. At early evolution times, such behaviour is attributed to a combination of the maximum defect counting resolution of a tangled configuration in a restricted (inhomogeneous) volume, and the quench rate exceeding internal system timescales, implying that the assumption of a well-defined local temperature starts to break down. The occurrence of a plateau for fast quenches has been also discussed in recent work in the context of $(2+1)$-dimensional holographic superfluids \cite{Chesler15}.   

{\em Defect Visualization and Nature:}
Typical experimental measurements made after an evolution time of about 250~ms and based on integration over different (radial or axial) directions after time-of-flight (TOF) expansion are shown in Fig.~\ref{fig_5} (left) for cases corresponding to different defect numbers. Such experimental images are in direct qualitative agreement with our numerically-generated ({\em in situ}) images at the correspondingly-long evolution times (Fig.~\ref{fig_5}, right); the latter numerically-generated results additionally enable direct visualization of the condensate phase (rightmost column), which is crucial for probing the nature of the emerging defects. From Fig.~\ref{fig_2} it is already evident that the initially-generated defects during the critical region are highly-excited, have random shapes, sizes and orientations. An interesting question is how such random defects evolve into the experimentally-observed solitonic vortices \cite{Donadello14}; these are vortices with squashed $2 \pi$ phase winding~\cite{Tylutki15} and vortex line lying in a transverse radial plane, which corresponds to the lowest energy configuration in a highly-elongated superfluid \cite{Brand02,Komineas03,Donadello14,Ku14}. Numerical analysis of the phase evolution  and corresponding density plots, appear to indicate a gradual evolution from the random distribution of different excited vortical states \cite{Mateo14}, into excited defects which are preferentially stretched along the transverse directions. Such defects gradually relax to solitonic vortices as the system equilibrates. (Supplementary Fig.~5 and \href{https://youtu.be/P4xGyr4dwKI}{Movie 4}). 
While it is impossible to define a precise time at which defects acquire a solitonic-vortex nature, due to the gradual nature of this process which also depends on quench rate, all our findings appear broadly consistent with a solitonic-vortex nature $\sim$100 ms after $t_{\rm bec}$. 
It is worth mentioning that a comparable timescale was found in \cite{Ku16} for the decay of phase-imprinted dark solitons into solitonic vortices in superfluid Fermi gases in a very similar elongated geometry.

\subsection{Coherence and Equilibration Dynamics}

Having demonstrated solid agreement with experimental observations in appropriate regimes, and the ability to further interpret those through our simulations, we now use our numerical scheme to provide a deeper insight into the complicated nonlinear dynamical evolution and equilibration of quenched systems, covering also regimes where no experimental measurements are currently available.

In our quenched dynamical evolution of an initially equilibrium thermal gas, we have seen the quenched system falling out of equilibrium around the critical region, and identified a subsequent time, $t_{\rm bec}$, associated with the onset of condensation, in a manner which enables direct comparison to experimental measurements.  Here we investigate the re-equilibration dynamics of such a system to a final state dictated uniquely by our final quench parameters ($\mu_{\rm final}$, $T_{\rm final}$).
We show that this relaxation process depends on $\tau_R$ in a nontrivial way: in particular, while the {\em condensate-number growth} dynamics depends solely on the growth timescale, $\tau_G$ (which is itself a function of $\tau_R$, see Fig.~\ref{fig_3}e), the {\em coherence growth}, is additionally sensitive to details of the defect-filled state of the system following the quench.
This points directly to the link between relaxation of quench-induced defects on the one hand and coherence growth and final system equilibration on the other.

From the condensate-number growth fits, we have identified two distinct dynamical regimes: for slow enough quenches ($\tau_R \gtrsim 300$ ms), the growth timescale is a monotonically-increasing function of the quench duration, whereas faster quenches ($\tau_R \lesssim 300$ ms) all exhibit a similar number growth timescale (Fig.~\ref{fig_3}e).
Nonetheless such rapid quenches lead to a notable increase in the number of spontaneously-generated defects, whose subsequent (``coarse-graining'') dynamics is crucial for the evolution of coherence.

To study the growth of coherence, we follow the procedure of the Cambridge group \cite{Navon15} by numerically shifting the wavefunction by a fixed amount and autocorrelating this with the unshifted copy of itself. This method provides an estimate of the coherence length of the system, $l_{\rm coh}$. Due to geometrical considerations, we focus here on the axial coherence length, obtained by transversal integration (see {\it Methods}).  In all cases we find that the integrated coherence length only starts increasing noticeably about 30 ms after $t_{\rm bec}$, consistent with the end of the previously identified rapid defect decay stage (Fig.~\ref{fig_4}a). 
The amount of vorticity present in the system sets a maximum limit to the dynamical system coherence length.  This is to be expected, and has already been noted, for example, in 1D \cite{Witkowska11,Karpiuk12} and 2D \cite{Jelic11}.
Importantly, however, we see that the coherence length of the slowest quenches grows much more rapidly and saturates at higher values. 
This observation points to the importance of a system expelling practically all of its defects before it can acquire a coherence length comparable to the system size.

\begin{figure*}[t!]
\includegraphics[width=0.8\linewidth]{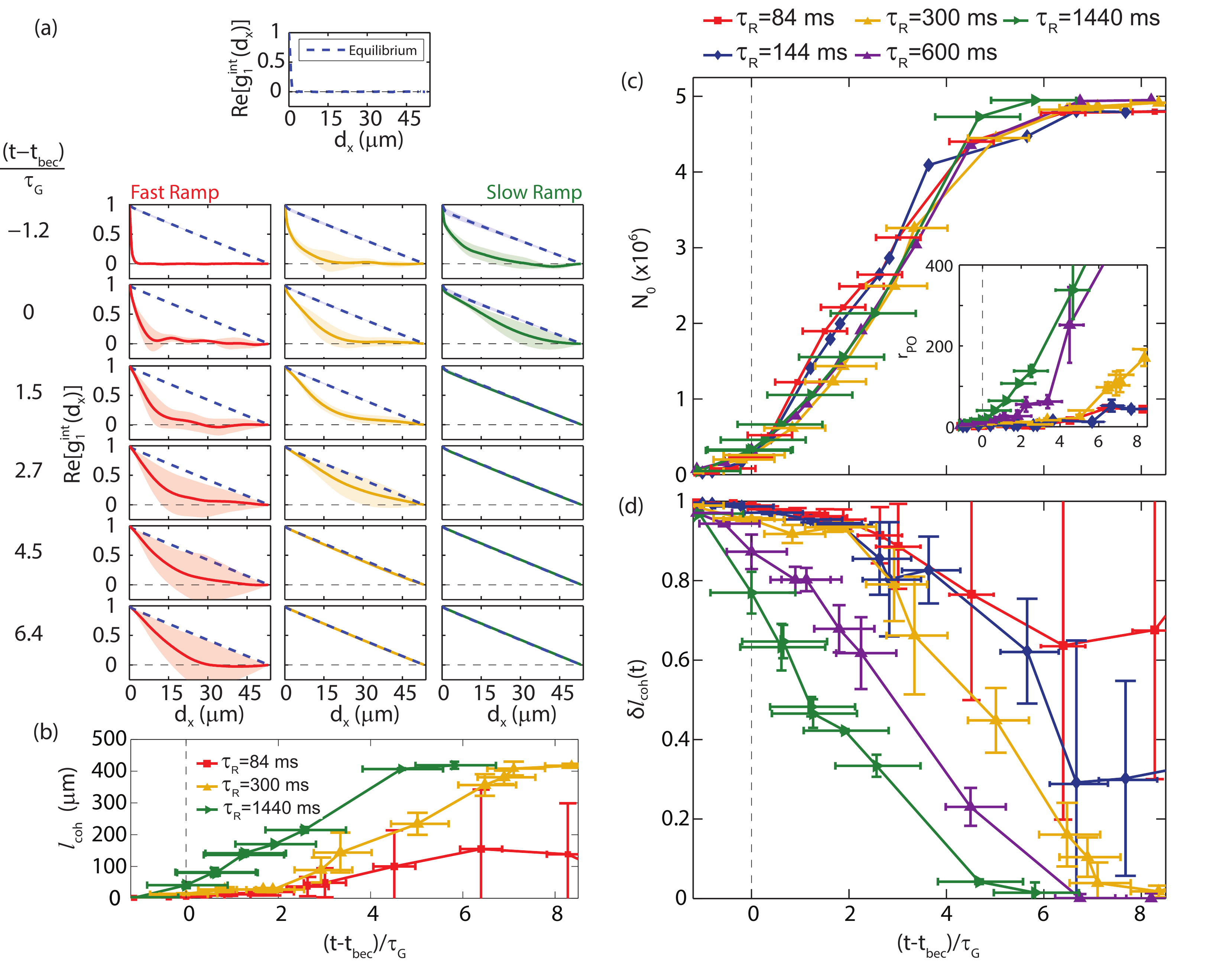}
\centering
\caption{Decoupling of Number and Coherence Growth.  
\textbf{a}  Evolution of the integrated first-order correlation function (Eq.~(7)) for fast ($\tau_R = 84$~ms, left), intermediate ($\tau_R = 300$~ms, middle) and slow ($\tau_R=1440$~ms, right) quench rates.  The top panel shows the correlation functions of the initial thermal state, when dynamical and equilibrium coherence lengths overlap. Subsequent images depict the dynamical values at time $t$ (solid lines/bands) and corresponding equilibrium values for the same ($\mu(t), T(t)$) combinations (dashed lines). For our chosen parameters, the equilibrium system is already in the phase-coherent condensate regime, for which $g_1^{int} \approx (1-|d_x|/L)$, for all times $t \gtrsim t_{\rm bec}$, with the correlation function for slower ramps approaching the corresponding equilibrium ones faster than for fast ramps.  
\textbf{b} Growth of the coherence length in time:
systems with more than 2 vortices on average exhibit coherence smaller than the system size, indicating that for the fastest ramps the system is still in the phase-fluctuating regime.  
\textbf{c} Condensate growth curves collapsing onto a single curve when time from the condensate onset is scaled to the intrinsic system growth timescale $\tau_G$. Inset shows corresponding ratio, $r_{PO}$, of most-populated to next-most-populated eigenmode of the one-body density matrix. 
\textbf{d} Corresponding scaled deviation, $\delta l_{\rm coh}(t)$, of dynamical to equilibrium coherence length [based on Eq.~(\ref{delta-l})]. At $t = t_{\rm bec}$, the slower ramps ($600$~ms, $1440$~ms) are closer to equilibrium than the faster ones. Slower ramps, scaled to their respective $\tau_{G}$, re-equilibrate faster to a phase-coherent BEC (for which $\delta l_{\rm coh} = 0$), with faster ramps only reaching full system coherence asymptotically. 
Statistical errors are shown by coloured bands in (a) and vertical error bars in (b) and (d). Horizontal error bars in (b)-(d) arise from the combination of the uncertainty in identifying $t_{\rm bec}$ and the fitting error for $\tau_G$. 
}
\label{fig_6}
\end{figure*}

Starting from the same initial thermal condition, Fig.~\ref{fig_6}a shows how the dynamical correlation functions (solid lines) evolve in cases of fast (left), intermediate (middle) and slow (right) ramps and compared to the corresponding  equilibrium functions (dashed lines).
Our analysis here focuses on the re-equilibration dynamics, and so all dynamical data presented here correspond to times $t>0$ (with $\mu(t)>0$) after the system has crossed the {\em ideal-gas} transition temperature, with times scaled to the quenched growth timescale $\tau_G$ in order to suppress corresponding differences in condensate number growth dynamics. In the context of our adopted definition for the correlation function \cite{Navon15}, this implies that the correlation function approaches a diagonal straight line as the coherence length approaches/exceeds the system size, which is always the case for the equilibrium systems considered here, due to their large atom numbers: this is decoupled from the fact that the equilibrium coherence length is increasing in absolute terms as the condensate size grows.
Looking at times $0 < t < t_{\rm bec}$ we see that although the corresponding equilibrium correlation functions already exhibit near-perfect coherence across the probed central region, the corresponding dynamical behaviour at $(t-t_{\rm bec})/\tau_G \approx -1.2$ deviates noticeably: in the case of the faster ramp (left column), the dynamical correlation function has only mildly increased from its incoherent thermal state initial value, contrary to a much stronger increase for the slower (quasi-adiabatic) ramp, which nonetheless also exhibits a phase coherence length significantly smaller than the quantum-degenerate system size. 
Fig.~\ref{fig_6}b shows the dynamical coherence length as a function of the rescaled timescale $(t-t_\mathrm{bec})/\tau_G$.
For all ramps the coherence length increases with time, but it does so in a much slower manner for the fast quench. As a result, significant coherence has already built up for the slower ramp already at our identified condensation onset time, $t_{\rm bec}$, highlighting that such a system approaches the equilibrium state rapidly in the early stages after crossing the phase-transition, even before significant number growth takes place; this is in stark contrast to the faster ramp, which still exhibits hardly any coherence.
Subsequent images highlight the emerging difference in the dynamical phase correlation function, $g_1^{int}(d_x)$ [defined in Eq.~(\ref{g1-int})],  in a most pronounced way, with the slow ramp becoming phase-coherent already for $(t-t_{\rm bec})/\tau_G \gtrsim 1.5$, when the system is still only at the initial part of its number growth curve, as opposed to the faster ramp whose coherence length is smaller than the system size even at the much later time $(t-t_{\rm bec})/\tau_G=12.7$. 

To quantify such re-equilibration dynamics further after the system has fallen out of equilibrium upon entering the critical phase-transition region, we introduce here the scaled deviation, $\delta l_{\rm coh}$,  of the dynamical correlation length $l_{\rm coh}^{\rm dyn}(t)$ from the corresponding equilibrium one, $l_{\rm coh}^{\rm equil}(\mu(t),\,T(t))$, defined by
\begin{equation}
\delta l_{\rm coh}(t) = \frac{l_{\rm coh}^{\rm equil}(\mu(t),\,T(t)) - l_{\rm coh}^{\rm dyn}(t) }{l_{\rm coh}^{\rm equil}(\mu(t),\,T(t))} \;.
\label{delta-l}
\end{equation}
Despite similar condensate growth rates (Fig.~\ref{fig_6}c) the corresponding coherence growth dynamics shown in Fig.~\ref{fig_6}d exhibit starkly distinct features, remaining strongly-dependent on the quench rate: in such relative timescales, slower ramps lead to much more rapid equilibration than the faster ramps;  the latter are slowed down by the detrimental role of the defects persisting within the system. Contrary to this, slow ramps which perturb the system less lead to the emergence of a nearly-defect-free and therefore phase-coherent condensate already around $t \approx t_{\rm bec}$.
The inset in Fig.~\ref{fig_6}c highlights the rapid emergence of a single macroscopically-occupied mode for the fastest ramps ($600$ and $1440$~ms), consistent with the rapid monotonic decrease of $\delta l_{\rm coh}(t)$, indicating the rapid crossover to a phase-coherent condensate.
The vertical error bars arising solely from our numerical averaging are significantly larger in the case of fast quenches, which is a measure of the deviation between different trajectories for the same ramp. 
This is easy to understand, since
the more vortices there are in the system, the more likely their configuration is to be significantly different from shot to shot.

\section{Discussion}

Motivated by recent experiments with dilute ultracold atomic gases, we have investigated numerically the dynamics of an equilibrium thermal gas quenched over a finite timescale to deep in the phase-coherent condensate regime.
Monitoring the entire evolution, we have presented an insightful graphical representation of the critical region dynamics, during which the dynamical system falls out of equilibrium with the corresponding-parameter equilibrium system,
through the dynamical symmetry-breaking spontaneous emergence of defects.
The emphasis of our analysis has been on the less-studied re-equilibration dynamics, addressing the interplay between defect emergence and dynamical evolution, and growth of coherence.

Depending on the quench duration we have identified different emerging dynamical regimes: for fast quenches, we observed a saturation in the number of detectable defects, associated with detrimental defect interactions and the inherent difficulty in counting randomly-oriented defects within a very tight volume. 
Rescaling the condensate number growth for all quench rates by the characteristic growth timescale for each ramp, we demonstrated the decoupling of coherence and number growth dynamics arising from the detrimental effect of defect emergence and propagation on system coherence.
Although the overall growth timescale might be longer for systems undergoing slower external cooling quenches, in such cases the dynamical system quickly re-approaches the corresponding equilibrium configuration, as the slow evolution enables it to mimic local equilibration during its growth, falling out of equilibrium only in a relatively small time window upon entering the region of critical fluctuations. In cases where the quench induces numerous defects, we have observed, as expected, enhanced shot-to-shot variability.

Our numerical analysis has also shed more light into the dynamical crossover from quasicondensation to true Bose-Einstein condensation, studied here in the context of an elongated geometry; such a geometry is known to lead to an enhanced decoupling of characteristic temperatures for the onset of density and phase fluctuations \cite{Petrov01,Proukakis06}, with such decoupling effectively translating into different growth rates for the phase-coherent and density-coherent parts of the system, and so different growth rates for coherence and quasicondensation.
Quasicondensation here refers to a defect-filled phase-fluctuating state with 
a coherence length smaller than the quantum-degenerate system size exhibiting suppressed density fluctuations
and spanning many largely-populated modes; this is directly contrasted to ``true'' condensation, which refers to a phase-coherent condensate with little, or no, defects and a single emerging macroscopically-occupied mode,
a definition which holds even if the system is growing in time.
Even though the quasicondensate stage in the critical region pre-empts all phase-coherent condensate growth due to the decoupling of the two corresponding characteristic temperatures, the dynamical quasicondensate regime is enhanced both in parameter space and in the temporal domain by the prolonged survival of the defects, thus being critically dependent on the quench rate responsible for their initial spontaneous generation.
For slow quenches and large enough atom numbers considered here, the system may already be largely phase-coherent, i.e., a ``true'' condensate, as soon as it has grown to a size which makes it experimentally detectable.
We note however that such findings are sensitive to the details of the system, and specifically here to both the system geometry and the chosen final state of the system, with reduced dimensionality enhancing such a distinction by enhancing the role of phase fluctuations \cite{Petrov01,Dettmer01,Richard03}.

In conclusion, we have  presented a systematic numerical study of the re-equilibration dynamics of a quenched  elongated 3D ultracold gas, 
demonstrating the decoupling of number and coherence growth in quenched quantum gases. Such decoupling is a consequence of the emergence and dynamics of the spontaneously-generated defects as the system crosses the phase transition. The detailed numerical visualization of the defect-filled quenched phase-transition dynamics allowed access to a broad temporal range of dynamics not typically experimentally-accessible, demonstrating the dominant role of defect interactions and decay in the early stages of condensate formation.
Moreover, our findings have been shown to be fully consistent with experimental measurements in the appropriate limits (and for the relevant quantities) where those exist. 

Our work is expected to be of relevance to a broad range of future investigations with quantum-degenerate systems, and could also have technological implications for dynamical control and state-engineering of a quantum system. Given that our numerical scheme has demonstrated good qualitative description also of the much-harder-to-model approach to the phase transition, we believe that our method could in the future offer further insight into delicate features of non-equilibrium condensate dynamics, including a critical assessment and extension of the inhomogeneous Kibble-Zurek phenomenon.

\section{Methods}

\noindent {\bf Experiments.}
We produce ultracold samples of sodium atoms in the internal state $|F,m_{\mathrm{F}}\rangle=|1,-1 \rangle$ in a cigar-shaped harmonic magnetic trap with trap frequencies $\omega_x/2 \pi=13$~Hz and $\omega_\perp/2 \pi=131.4$~Hz.  The thermal gas is cooled down via forced evaporative cooling and pure BECs of typically $10^7$ atoms are produced. The part of the evaporation ramp in the vicinity of the transition is performed at different rates, from  $50$~kHz/s to $2$~MHz/s. The quench ramp is followed by a variable wait time, during which a radio frequency shield is kept on to prevent from heating. After that, the atoms are released from the trap and are observed in two possible ways: Either we take simultaneous absorption images of the full atomic distribution along the radial and the axial directions \cite{Donadello14,Donadello16} or we extract, uniformly, a small amount of atoms from the trapped sample and image it after a short time of flight \cite{Serafini15,Serafini17}.
The protocol is such that images are taken, and vortices are counted, after a fixed overall time interval from the BEC transition point, which is clearly identifiable for each quench ramp.  We are also able to precisely identify the frequency $\nu$ of the RF field at $T_{\mathrm{c}}$, as well as  to control the temperature variation in time, $(\partial T/\partial t)$, {\it via} the speed of the evaporation ramp $(\partial \nu/\partial t)$ \cite{Donadello16}. The defects that we observe at the time of imaging are quantized vortex lines which are seen as dark stripes when looking at the BEC from a radial direction after time-of-flight. The natural size of the defects in the trapped BEC, at the end of the cooling ramp, is of the order of the \textit{in-situ} healing length $\xi$, which is of the order of $200$~nm. After a long TOF, the defect size becomes larger than our imaging resolution of $3\ \mu$m. The presence of a levitating magnetic field gradient makes it possible to achieve long TOF preventing the BEC from falling. The measured vortex number is averaged over many experimental realizations in order to get good statistical samples for each experimental condition.

\bigskip

\noindent {\bf Numerical Model.}
Out study is performed by means of the (simple growth) stochastic projected Gross-Pitaevskii equation \cite{Blakie08} (see also related model without projector \cite{Stoof01,Proukakis08,Cockburn10,Proukakis13}), already demonstrated as a useful tool for the quenched crossing of the BEC phase transition \cite{Weiler08,Su13,Liu16}. In brief we simulate the low-lying highly-occupied modes of the system, denoted by the classical field (or c-field \cite{Blakie08}) $\Psi_\mathcal{C}(\mathbf{r},t)$  through the dynamical equation
\begin{equation}
d\Psi_\mathcal{C}(\mathbf{r},t)=\mathcal{P}_\mathcal{C}\Big\{-\frac{i}{\hbar}\mathcal{L}+\frac{\gamma}{\hbar}[\mu(t)-\mathcal{L}]\Big\}\Psi_\mathcal{C}(\mathbf{r},t)dt+dW_\gamma(\mathbf{r},t),
\end{equation}
where $\mathcal{P}_{\mathcal{C}}$ is the projection operator truncating the modes above the c-field regime (so above an appropriately-identified energy cutoff), 
$\gamma$ is a constant determining the condensate growth timescale, $\mu(t)$ is the time-dependent chemical potential,
and $\mathcal{L}=-\hbar^2 \nabla^2/2m+V(\mathbf{r})+g|\Psi_\mathcal{C}(\mathbf{r},t)|^2$
with $V(\mathbf{r})=(m/2)\left[\omega_x^2x^2+\omega_\perp^2(y^2+z^2)\right]$. 
Fluctuations are included through the complex white noise, $dW_\gamma$, defined by
$\langle dW^\ast(\mathbf{r},t)dW(\mathbf{r}^\prime,t)\rangle=[2\gamma k_BT/\hbar]\delta_\mathcal{C}(\mathbf{r}-\mathbf{r}^\prime)$
where 
​$\delta_\mathcal{C}(\mathbf{r}-\mathbf{r}^\prime)=\sum_{n\in\mathcal{C}}\psi_n^\ast(\mathbf{r})\psi_n(\mathbf{r}^\prime)$
in the chosen orthogonal basis set, $\{\psi_n(\mathbf{r})\}$. 
The constant $\gamma$ can be analytically approximated (at least for near-equilibrium cases) as \cite{Penckwitt02,Proukakis08}
\begin{eqnarray}
\gamma & \approx & 2 \left( \frac{m}{2 \pi \hbar^2} \right)^3 \frac{g^2}{k_B T} \int dE_2 \int dE_3 (1+N_1)N_2 N_3  \nonumber \\ 
& \approx & {\rm few} \times \frac{4 m k_B T}{\pi \hbar^2} a^2
\end{eqnarray}
where $g=4 \pi \hbar^2 a /m$ is the interaction strength, corresponding to the $s$-wave scattering length $a$. Although the above formula indicates the leading functional dependence of $\gamma$, the factor of ``few'' conceals within it the fact that this is an effective scaling, and so one should only rely on this for order-of-magnitude estimates.
Consistent with earlier related analysis \cite{Weiler08}, here we treat $\gamma$ as a fitting parameter. Comparing to experimental condensate growth data for $dT/dt=5.6$~$\mu$K/s (Fig.~\ref{fig_3}b), we identify  an ``optimal'' value $\gamma =0.005$, which is about 10 times the analytically-predicted value for the initial temperature. 
To mimic the experimental cooling process, the chemical potential $\mu(t)$ and temperature $T(t)$ are quenched linearly within a ramp duration $\tau_R$. The initial and final values of $(\mu,T)$ are set as $(-22\hbar\omega_\perp,790$ nK $)$ and $(22\hbar\omega_\perp,210$ nK $)$, corresponding to a change of total equilibrium atom number (when also including above cut-off atoms under the usual assumption that they are static) from $22\times10^6$ to $6.6\times10^6$ with a 75\% final condensate fraction. In our simulations, which start from a highly incoherent equilibrium state well above $T_c$, the time $t=0$ is chosen as the time when $\mu(t)=0$, since most interesting dynamics occurs after this time, implying that in any given simulation, the initial (equilibrium) configuration is at a time $t = -\tau_R/2$.\\

We solve the SPGPE with 4th-order Runge-Kutta in a plane-wave basis using a grid size $L_x=54a_{ho,x}$ along the x and $L_y=L_z=6a_{ho,x}$ along the transverse directions, where $a_{ho,x}=\sqrt{\hbar/m\omega_x} \approx 5.8\mu$m is the characteristic harmonic oscillator length in the long direction (x-axis); we use a temporal discretization $dt=10^{-3} /\omega_{x}$ and an energy cutoff fixed at 2.5 times the value of the final chemical potential ($22\hbar\omega_{\perp}$) in a grid consisting of $N_x=1170$ and $N_y=N_z=130$ points.
Simulations are run on Newcastle University's High-Performance-Computing cluster, Topsy, using $20$ to $24$ nodes. A single dynamical run takes between $(120 - 300)$ CPU hours with an additional $\approx 40$ CPU hours for the Penrose-Onsager diagonalization of the selected snapshots. We estimate the total amount of presented simulations to have taken over $10,000$ CPU hours.

\bigskip

\noindent {\bf Identification of the Condensate.}
The one-body density matrix is defined as
\begin{equation}
\begin{array}{rl}
\rho(\mathbf{r},\mathbf{r}^\prime;t)\equiv &\overline{\langle\Psi_\mathcal{C}^*(\mathbf{r},t)\Psi_\mathcal{C}(\mathbf{r}^\prime,t)\rangle}\nonumber \\
\\
=&\displaystyle\nonumber \sum_{j=0}^{N_{\rm sample}-1}\frac{\Psi^\ast_\mathcal{C}\left(\mathbf{r},t+j\delta t\right)\Psi_\mathcal{C}\left(\mathbf{r}^\prime,t+j\delta t\right)}{N_{\rm sample}} \nonumber 
\end{array}
\end{equation}
where $N_{\rm sample}$ is the sample number of the subensemble average and $\delta t$ is set as $\Delta t/N_{\rm sample}$ with  an appropriately short time-interval $\Delta t$ (so that the system dynamics is not masked). Such short-time averaging mimics the ensemble averaging based on the ergodicity hypothesis \cite{Blakie08}. The notation $\overline{\langle...\rangle}$ used in this and the next subsections denotes the short-time subensemble average. In our simulations, $N_{\rm sample}=101$, and the tests of probing $\Delta t$ provide a value of around $8$~ms for our simulations, which is smaller than the characteristic time scale of the harmonic trap $\tau_{ho}=1/\omega_x\approx 12.2$~ms.  The condensate, or Penrose-Onsager (PO) mode~\cite{Penrose56} at a given time $t$ is identified as the eigenmode of the one-body density matrix  $\rho(\mathbf{r},\mathbf{r}^\prime;t)$ with the largest eigenvalue.
To assess the degree of fragmentation of the condensate, in the sense of competition between different highly-occupied modes, we evaluate the ratio, $r_{PO}$,  of the largest to the second largest eigenvalues of $\rho(\mathbf{r},\mathbf{r}^\prime;t)$.  \\

\bigskip

\noindent {\bf Correlation Function Analysis.}
We follow the procedure of the Cambridge quenched-dynamics experiment \cite{Navon15}, which measured the correlation function by interefering a displaced copy of the system with itself.
Specifically we define the function
\begin{equation}
g_{1}^{H}(d_x,t)=\iint dydz\int_0^L dx \Psi^\dagger(\mathbf{r},t)\Psi(\mathbf{r}+d_x\hat{x},t)
\end{equation}
where $L$ is a chosen length. Following Ref.~\cite{Navon15}, the correlation length $l_{\rm coh}$ is extracted from $g_1^H$ by fitting it with $(1-|d_x|/L)\exp(-{|d_x|/l_{\rm coh}})$, in which the triangular-shape function in the bracket arises from the integration. 
Here we probe the spatial coherence of our c-field wavefunction $\Psi_\mathcal{C}$ within a region $[-L/2,\;L/2]$ with $L\sim54$ $\mu$m, and numerically evaluate this through the phase-phase correlation function by
\begin{equation}
g_1(d_x,y,z;t)=\overline{\left\langle\int_{-L/2}^{L/2} dxH[e^{i\phi(\mathbf{r},t)}]\right\rangle},
\end{equation}
where the operation $H[f]$ is defined as
\begin{equation}
H[f](d_x)\equiv\int_{-L/2}^{L/2}dxf^\ast(\mathbf{r})f(\mathbf{r}+d_{x}\hat{\mathbf{x}}),
\end{equation}
and $\phi(\mathbf{r},t)$ is the argument of $\Psi_\mathcal{C}(\mathbf{r},t)$.
To obtain transversal averaging we use the integrated version,
\begin{equation}
g_{1}^{int}(d_x,t)=\iint^\prime dydz\Big[w_{PO}(d_x,y,z;t) g_1^H(d_x,y,z;t)\Big],
\label{g1-int}
\end{equation}
where the prime integration is performed over the yellow region (i.e. density isosurface at values of 0.1\% of the final c-field density.)
To take finite-size effects into account we have also introduced into the above definition a density-dependent weighting function $w_{PO}(d_x,y,z;t)$ which assigns higher weighting to the large-density regions. This is defined here through
$w_{PO}(d_x,y,z;t)=H[n_{PO}(\mathbf{r})/n_{PO,\textrm{peak}}]/\mathcal{N}$, where $\mathcal{N}$ ensures the normalization  condition $\iint^\prime dydzw_{PO}(d_x,y,z;t)=1$ is satisfied at any given $d_x$. (The second-order correlation function shown in Supplementary Fig. 5 and \href{https://youtu.be/P4xGyr4dwKI}{Movie 4} is defined as the onsite correlation function $g_2(\mathbf{r})\equiv g_2(\mathbf{r},\mathbf{r},\mathbf{r},\mathbf{r})=\overline{\langle|\Psi_{\mathcal{C}}(\mathbf{r})|^4\rangle}/[\overline{\langle|\Psi_{\mathcal{C}}(\mathbf{r})|^2\rangle}]^2$.)

\bigskip

\noindent {\bf Dynamical Timescales.}
Consistent with typical experimental measurements, in which $T_{c}$ is identified as the time of emergence of an observable condensate, we define here the ``onset'' or ``delay'' time for condensate growth, $t_{\rm bec}$, as the moment that the number of atoms in the condensate,  $N_0$, reaches 5\% of the final total particle number (including in our considerations the particle number above the c-field region, which is assumed to be static).
We have {\em a posteriori} verified this to provide (when used with our value of  $\gamma$) an excellent description of condensate growth across all experimentally-probed regimes, and have also checked that the main findings presented in this paper are insensitive to the details of such definition. \\
\noindent In addition to  $t_{\rm bec}$, we also define the condensate growth timescale, $\tau_G$, which is  extracted by fitting the condensate growth curve over the entire temporal range $t$ with
\begin{equation}
N_0(t)=N_{0,i}+\frac{N_{0,f}-N_{0,i}}{1+\exp[-(t-\tilde{t} \, )/\tau_G]} \;,
\label{eq:sshape}
\end{equation}
where $N_{0,i}$ ($N_{0,f}$) denote the initial (final) PO condensate atom numbers, $\tilde{t}$ is the moment that $N_0(\tilde{t}\, )$ reaches the mid-value $(N_{0,i}+N_{0,f})/2$, and $\tau_G$ is the single fitting parameter. We note here two things: firstly, that the mid-value is unique to all numerical growth curves, since we have a unique set of experimentally-relevant initial and final parameters in our simulations; moreover, we have checked that the extracted $\tau_G$ values are largely insensitive to whether the fit is performed over the entire temporal range $t$, or whether it is constrained to values $t \ge t_{\rm bec}$, suggesting the independence of the two timescales $t_{\rm bec}$ and $\tau_G$.

\bigskip

\noindent {\bf Defect Identification.}
In our work we identify the location of vortices by the regions of high velocity, $\mathbf{v}(\mathbf{r})=(\hbar/m)\textrm{Im}[\Psi_{\mathcal{C}}^\ast(\mathbf{r})\nabla\Psi_{\mathcal{C}}(\mathbf{r})]/|\Psi_{\mathcal{C}}(\mathbf{r})|^2$, characterizing the region around the vortex core. By scanning the whole local maximum of the velocity field within the yellow region, we identify the positions of the vortex cores. 

\bigskip

\noindent{\bf Statistical Analysis and Error Bars.}
For each numerically-simulated quench rate, we have analysed between $\cal{N}=$ 3 and 7 independent-noise realisations. 
The statistical uncertainty in the vortex number, $N_v$, was estimated as 
\begin{equation}
\Delta N_v =\displaystyle\sqrt{\displaystyle\sum_{i=1}^{\cal N}\left[\frac{\left((N_v)_i-\langle N_v \rangle \right)^2}{{\cal N}^{2}}\right]+\frac{1}{\cal{N}}} \,\,\,\,.
\end{equation}
In counting the vortex number, $N_v$, we also estimated the possible systematic errors introduced by the use of subjective criteria in the identification of single vortices in situations where a vortex is at the boundary of the condensate or two vortex lines are very close to each other. However, we have checked that the corresponding uncertainty is significantly smaller than the statistical error defined above. \\
\noindent Our procedure for assigning errors to the determination of the characteristic timescales is as follows:\\
- the condensate onset time, $t_{\rm bec}$ was defined as the time at which the condensate (Penrose-Onsager) atom number reaches 5\% of the total final atom number. Error bars in our determination of $t_{\rm bec}$ arise from shifting the (heuristic) value of 5\% between 3\% and 7\%, values which are still consistent with the experimental growth curves reported in Fig.~3.\\
- the errors in the quenched growth timescales, $\tau_G$, arise from the 95\% confidence bounds of the fit to our numerical growth curves with Eq.~(8): the quality of the fits can be seen in Supplementary Fig.~4.\\
Those two errors are treated as independent in the determination of temporal error bars for the scaled time $(t-t_{\rm bec})/\tau_G$ discussed in Fig.~6.

\noindent {\bf Data Availability}
Data supporting this publication is openly available
under an ��Open Data Commons Open Database License��.
Additional metadata are available at: http://dx.doi.org/..................... Please contact Newcastle Research Data
Service at rdm@ncl.ac.uk for access instructions.

\noindent {\bf Acknowledgements}
N.P.P. and I.-K.L. acknowlelge discussions with J{\'e}r{\^o}me Beugnon, Paolo Comaron, Leticia Cugliandolo, Jean Dalibard, Zoran Hadzibabic, Fabrizio Larcher, Nir Navon and Rob Smith during the course of the project, computational assistance by Kean Loon Lee and George Stagg and help with diagonalizing huge matrices from T.-M. Huang and W.-W. Lin. We acknowledge financial support from the Taiwan MOST 103-2112-M-018-002-MY3 grant (I.-K.L. abd S-C.G.), the UK EPSRC Grant No. EP/K03250X/1 (I.-K.L., N.P.P. ), and from Provincia Autonoma di Trento.  
\\

\noindent {\bf Author Contributions}
I.-K.L. undertook all the numerical simulations and analysis, in direct consultation with N.P.P. who coordinated the research, led the interpretations and produced the first draft. S.D, G.L. and G.F. designed and conducted the experiments, which were analysed jointly with F.D. Theoretical aspects were discussed by I-K.L, S.-C.G., F.D and N.P.P. All authors contributed to discussions, final data analysis and interpretations, and the final form of the manuscript.\\

\noindent {\bf Competing financial interests} The authors declare no competing financial interests.\\

\noindent {\bf Materials \& Correspondence} All correspondence and material requests should be addressed to N.P.P.


\end{document}


\title{Supplementary Material: \\ Dynamical Equilibration Across a Quenched Phase Transition \\ in a Trapped Quantum Gas}

\author{I.-K. Liu$^{1,2}$, S. Donadello$^3$, G. Lamporesi$^3$, G. Ferrari$^3$, S.-C. Gou$^{2}$, F. Dalfovo$^3$, N.P. Proukakis$^1*$}
\maketitle


In this Supplementary material we provide more detailed examples of the evolution presented in this manuscript.

\begin{itemize}

\item {\bf Supplementary Fig. 1} shows side-by-side the evolution of characteristic density isosurfaces of the highest-populated (Penrose-Onsager) mode for a single run based on the same dynamical noise sequence and 3 different quench rates, characterised by their duration $\tau_R$.
The more tangled evolution at early times, and the longer survival of defects in the fastest quench is visible.

\item {\bf Supplementary Fig.~2} shows characteristic examples of evolution of the Penrose-Onsager mode for a given quench rate ($\tau_R = 144$~ms), demonstrating clearly the stochastic nature of the defect generation, and the stark differences between different numerical runs -- a detailed analysis of which sets the error bars in Fig.~4 of the main paper.

\item {\bf Supplementary Fig.~3} shows the evolution of the Penrose-Onsager mode in a focused spatial region and small time-intervals for a single-run corresponding to a fast quench ($\tau_R=18$ms), clearly demonstrating the various types of processes that dominate the early physics of the system as it crosses the phase transition after which the initially-tangled-up generated defects interact and relax.

\item {\bf Supplementary Fig.~4} demonstrates our fitting procedure for extracting the condensate quenched growth timescale $\tau_G$.

\item {\bf Supplementary Fig.~5} shows different projections of the condensate wavefunction, phase and local second-order correlation function, depicting clearly the defect relaxation to solitonic vortices at late evolution times.

\end{itemize}

\vspace{1.0cm}

We also provide here real-time evolution movies corresponding to the data shown in Supplementary Figures 1, 2, 3 and 5 (with more snapshots).

\begin{itemize}

\item \noindent {\em Caption to Movie Liu-et-al-\href{https://youtu.be/3q7-CvuBylg}{SM-Movie-1}:} \\
{\bf Dependence of Condensate Growth and Equilibration on Quench Rate:}
Evolution of typical density isosurfaces of the highest-occupied (Penrose-Onsager) modes for different ramp durations $\tau_R = 84,\, 144$ and 600~ms (from left to right). [See also Supplementary Material Fig. 1]

\item \noindent {\em Caption to Movie Liu-et-al-\href{https://youtu.be/-Gymaiv9rC0}{SM-Movie-2}:}\\ 
{\bf Shot-to-Shot Fluctuations During Condensate Growth:}
Shot-to-shot variations in the evolution of typical density isosurfaces of the highest-occupied (Penrose-Onsager) modes for a ramp duration $\tau_R = 144$ ms, with depicted images starting just before the system reaches the condensation onset time $t_{\rm bec}$. [See also Supplementary Material Fig. 2]

\item \noindent {\em Caption to Movie Liu-et-al-\href{https://youtu.be/w-O2SPiw3nE}{SM-Movie-3}:} \\
{\bf Emergence, Interaction and Relaxation of Vortices in a Growing Condensate:}
Evolution of typical density isosurfaces of the highest-occupied (Penrose-Onsager) mode for a fast ramp ($\tau_R = 18$ ms) clearly demonstrating the complexity of vortex tangle generation, unravelling and subsequent defect interaction and relaxation. [See also Supplementary Material Fig. 3.]

\item \noindent {\em Caption to Movie Liu-et-al-\href{https://youtu.be/P4xGyr4dwKI}{SM-Movie-4}:}\\
{\bf Dynamical Condensation Representation:} 
Evolution of the highest-occupied (Penrose-Onsager) mode for $\tau_R = 144$ ms, depicted from two different viewing angles for density profile of Penrose-Onsager mode, and corresponding sliced profiles showing the related information for the density, phase and local second-order correlation function on $y=0$ (top) and $z=0$ (bottom) planes. [See also Supplementary Material Fig. 5 which also gives corresponding definitions.]

\end{itemize}

\begin{figure*}[hb]
\includegraphics[width=1\linewidth]{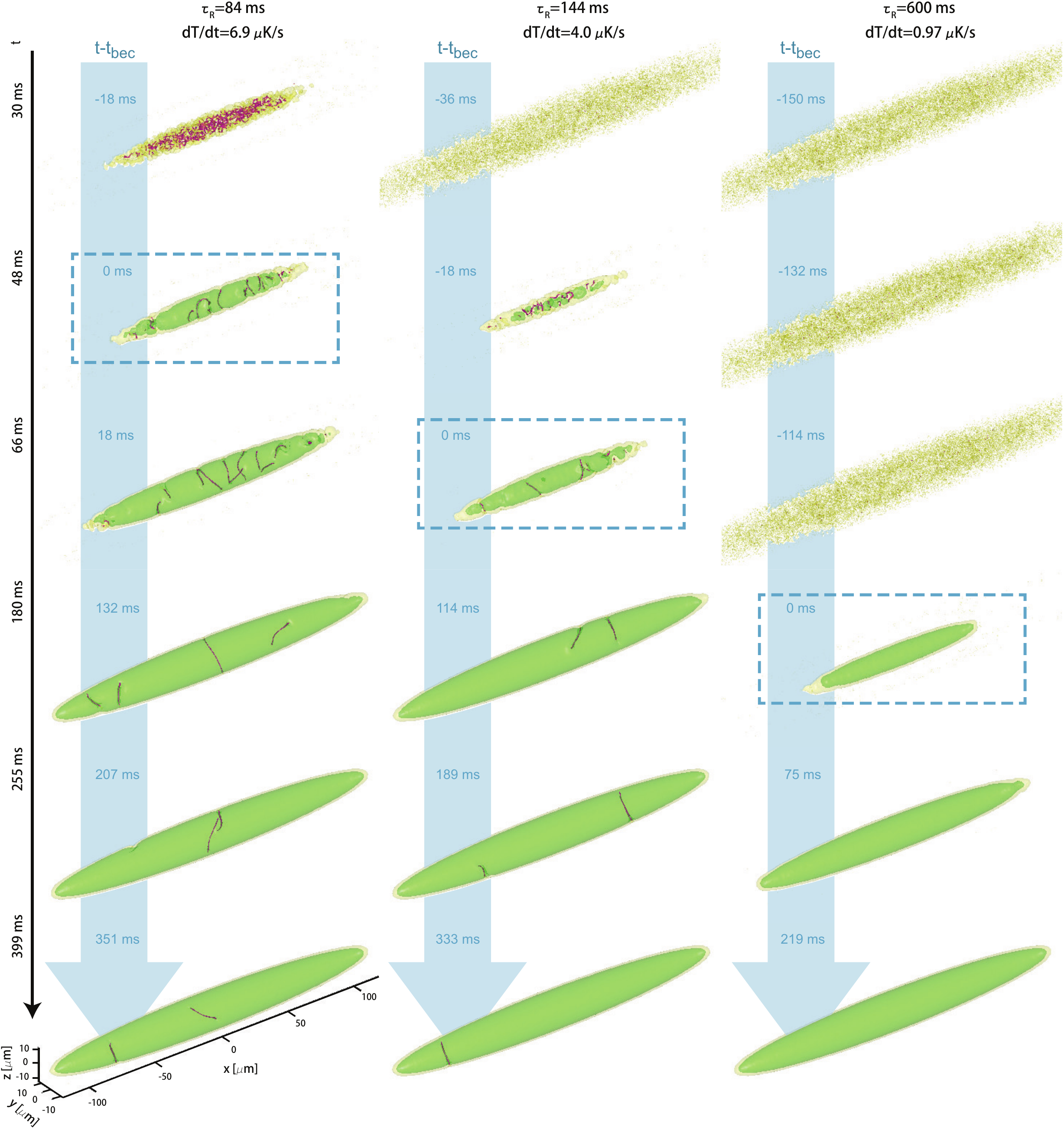}
\centering
\caption{
{\bf Dependence of Condensate Growth and Equilibration on Quench Rate:}
Shown are single-realisation density profile evolutions for 3 typical ramps (from left to right: $\tau_R = 84,\, 144,\, 600$ ms), with all subplots depicting density isosurfaces of the highest-occupied (Penrose-Onsager) modes at values 3\% (green regions) and 0.1\% (yellow regions). Each row corresponds to a different absolute time $t$ from the moment the system is quenched, also depicting the corresponding shifted time $(t-t_{\rm bec})$ which depends on $\tau_R$ (and so varies from column to column). All cases demonstrate an initial narrowing of the spatial distribution as the system approaches the critical region. Faster quenches are shown to have a more violent random defect tangle generation at $t \lesssim t_{\rm bec}$ thus exhibiting more defects at $t_{\rm bec}$ [boxed plots] when the condensates have similar sizes. At late times, faster quenches exhibit more vortices, with the slowest quench considered here never acquiring a vortex within the high density region. Runs shown here are based on exactly the same dynamical noise sequence and are for illustrative purposes.
Our estimated values for $t_{\rm bec}$ for the three ramps are respectively (from left to right): 
$t_{\rm bec}(\tau_R=84 \, {\rm ms}) = 48_{-3}^{+4}$ ms,  $t_{\rm bec}(\tau_R=144 \, {\rm ms}) = 66_{-3}^{+6}$ ms, and $t_{\rm bec}(\tau_R=600 \, {\rm ms}) = 180_{-12}^{+6}$ ms.
(See also \href{https://youtu.be/3q7-CvuBylg}{Suppementary Movie 1}).
}
\label{fig_1}
\end{figure*}

\begin{figure*}[hb]
\includegraphics[width=1\linewidth]{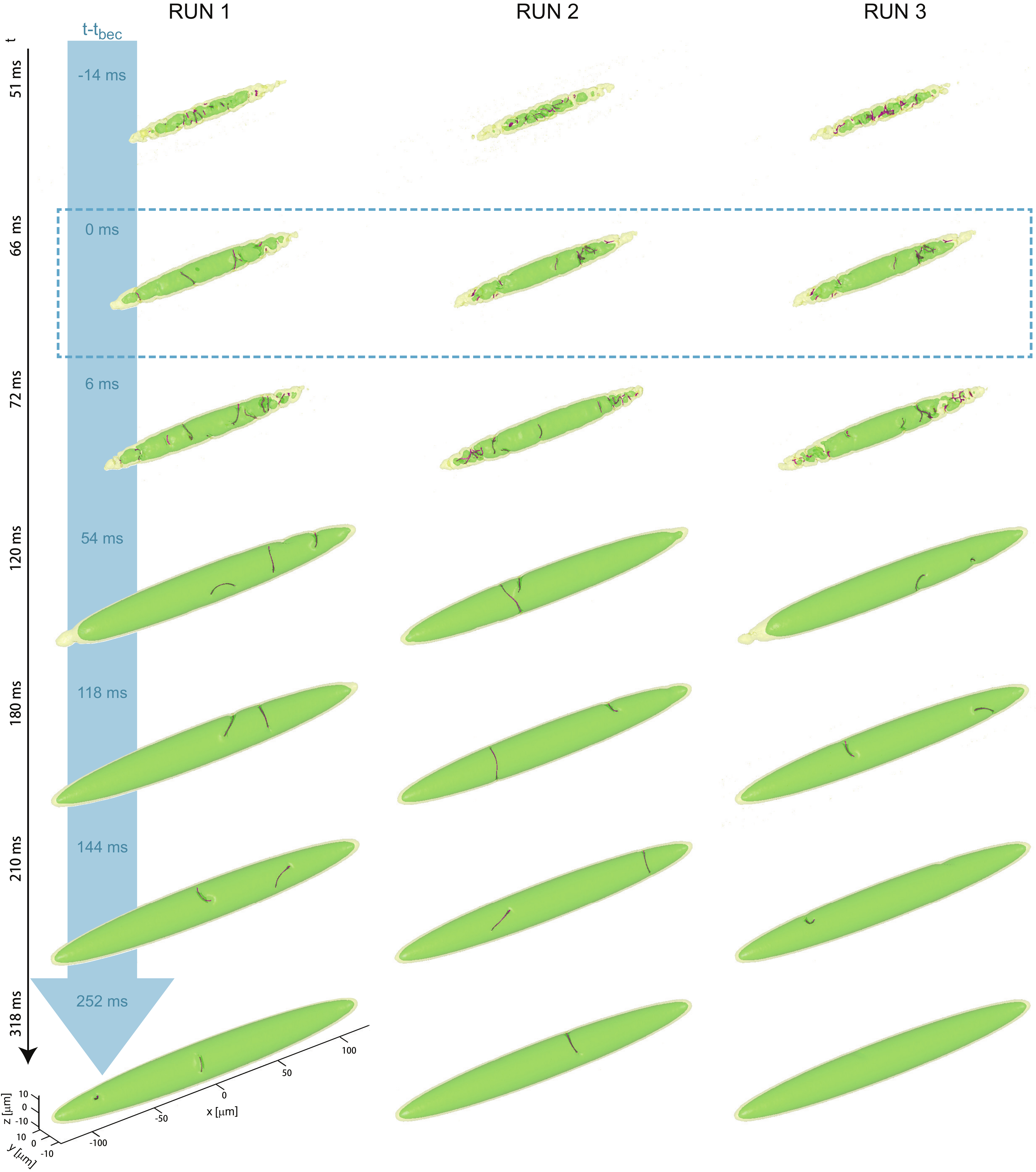}
\centering
\caption{
{\bf Shot-to-Shot Fluctuations During Condensate Growth:}
Shown are 3D visualizations of the characteristic density profiles of the highest-occupied (Penrose-Onsager) mode for 3 different numerical realisations for the $\tau_R=144$ ms ramp, which remains in place until $t=\tau_R/2=72$ms ($t-t_{\rm bec}=6$ms). The difference in defect number between equal-parameter runs at late-time evolutions sets the error bars in our vortex counting shown in the main text.
Our estimated value for the condensate onset time $t_{\rm bec}(\tau_R=144 \, {\rm ms}) = 66_{-3}^{+6}$ ms.
(See also \href{https://youtu.be/-Gymaiv9rC0}{Suppementary Movie 2}).
}
\label{fig_2}
\end{figure*}

\begin{figure*}[hb]
\includegraphics[width=1\linewidth]{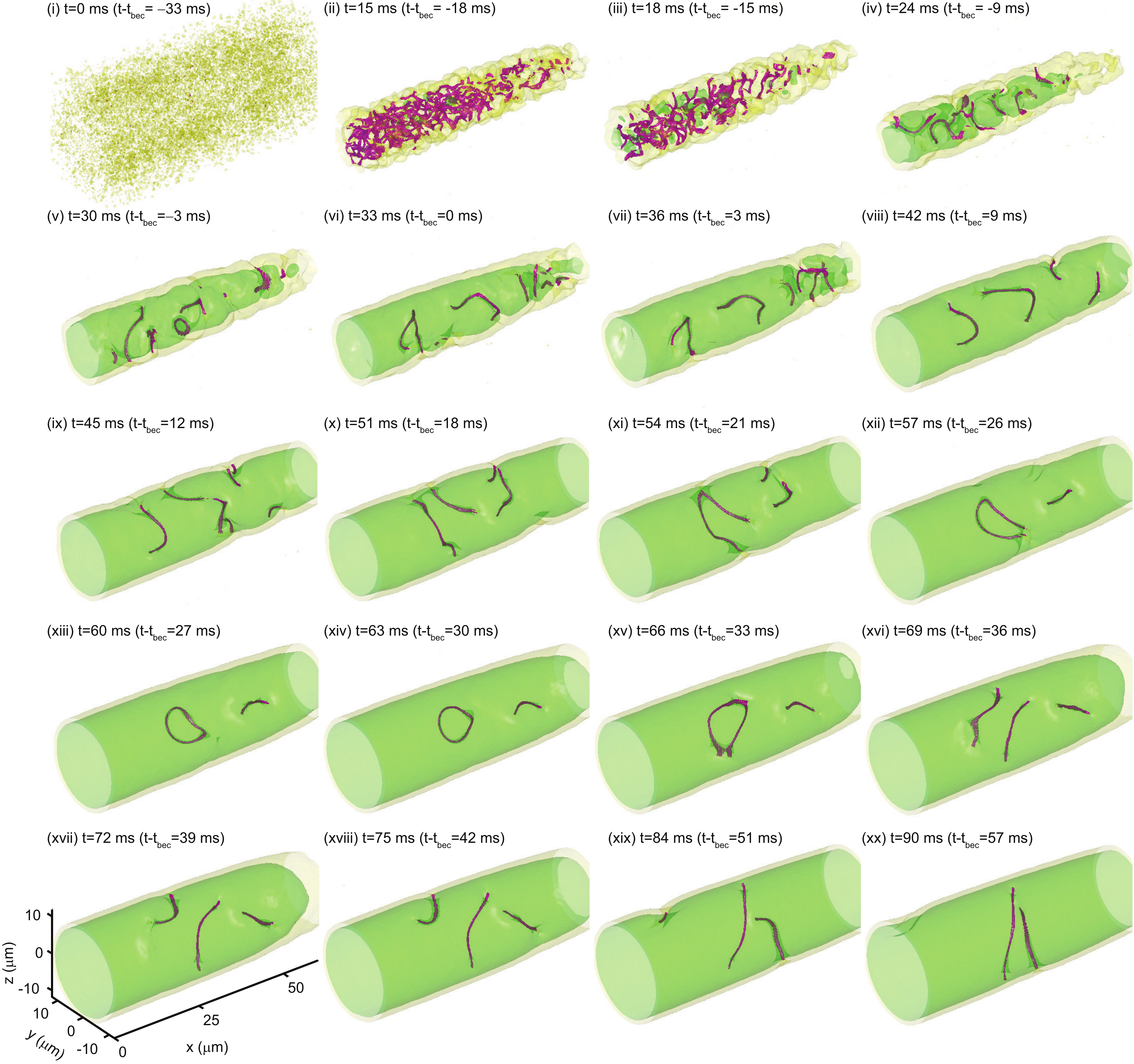}
\centering
\caption{
{\bf Emergence and Interaction of Vortices in a Growing (Quasi)condensate:}
Consecutive images depict gradual dynamical evolution of the density profiles of the highest-occupied (Penrose-Onsager) mode as the phase transition is crossed violently due to a rapid external quench ($\tau_R=18$ ms).
Our focussed region, spanning only a subset of our numerical grid and of the final system size, depicts clearly both the initial defect tangle complexity (15 - 18 ms), tangle unravelling with associated subsequent increase in mean distances between defects (24 - 45 ms), occurring alongside  defect interactions, including reconnections (with a clear isolated example show e.g.  during 45 - 54 ms), vortex ejection (top right part of plot for 54 - 57 ms, and top left for 84 - 90 ms), dynamical vortex ring generation as intermediate defect collisional stage (63 ms), with clear evidence of transversal vortex stretching (e.g. 51 ms, 90 ms).
Our estimated value for the condensate onset time $t_{\rm bec}(\tau_R=18 \, {\rm ms}) = 33_{-3}^{+3}$ ms.
(See also \href{https://youtu.be/w-O2SPiw3nE}{Suppementary Movie 3} demonstrating evolution over entire condensate length and up to later times).
}
\label{fig_3}
\end{figure*}

\begin{figure*}[hb]
\includegraphics[width=1\linewidth]{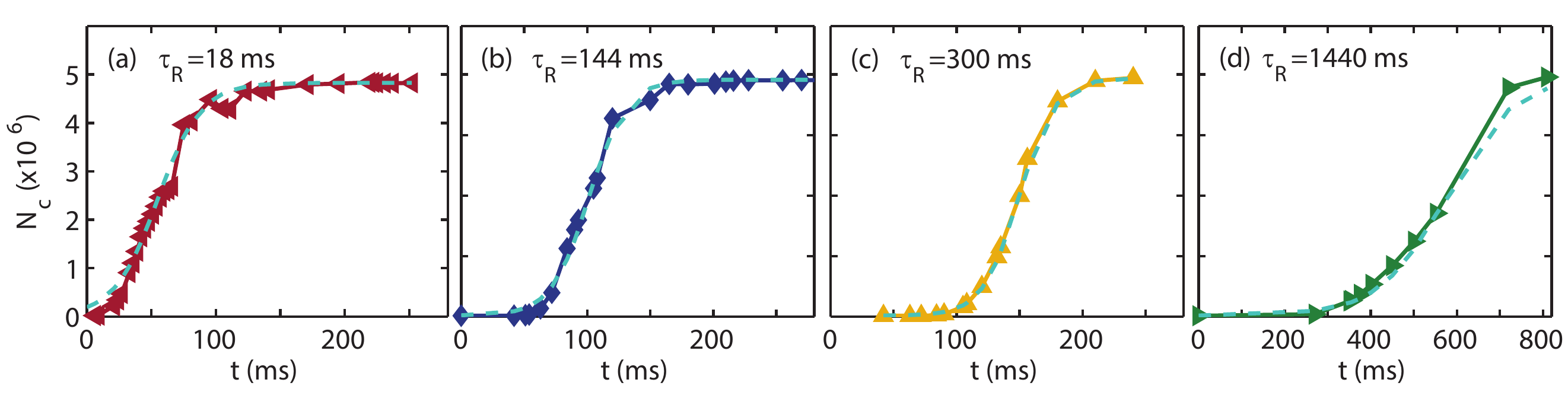}
\centering
\caption{
{\bf Condensate Growth and Emerging Timescales:}
Typical condensate growth curves as a function of (absolute) time $t$, averaged over $3$ realizations, shown from the moment that $\mu=0$, alongside corresponding fits (dashed cyan lines) with Eq.~(8) from {\em Methods} for ramps with (from left to right) $\tau_R = 18, \, 144, \, 300, \, {\rm and} \, 1440$ ms. The errors in the Penrose-Onsager number are smaller than the size of the depicted points, reconfirming the consistency of the Penrose-Onsager analysis despite the different density profiles between runs with same $\tau_R$. Corresponding extracted values of  $\tau_G$ are $16\pm2$~, $14.9\pm1.7$,$14.3\pm1.0$ and $78\pm13$~ms, with errors representing the 95\% confidence level of the fits.
For reference, our estimated values for $t_{\rm bec}$ for the ramps depicted here are respectively (from left to right): 
$t_{\rm bec}(\tau_R=18 \, {\rm ms}) = 33_{-3}^{+3}$ ms,  $t_{\rm bec}(\tau_R=144 \, {\rm ms}) = 66_{-3}^{+6}$ ms, $t_{\rm bec}(\tau_R=300 \, {\rm ms}) = 108_{-3}^{+7}$ ms, and $t_{\rm bec}(\tau_R=1440 \, {\rm ms}) = 354_{-39}^{+21}$ ms.
}
\label{fig_4}
\end{figure*}

\begin{figure*}[hb]
\includegraphics[width=1\linewidth]{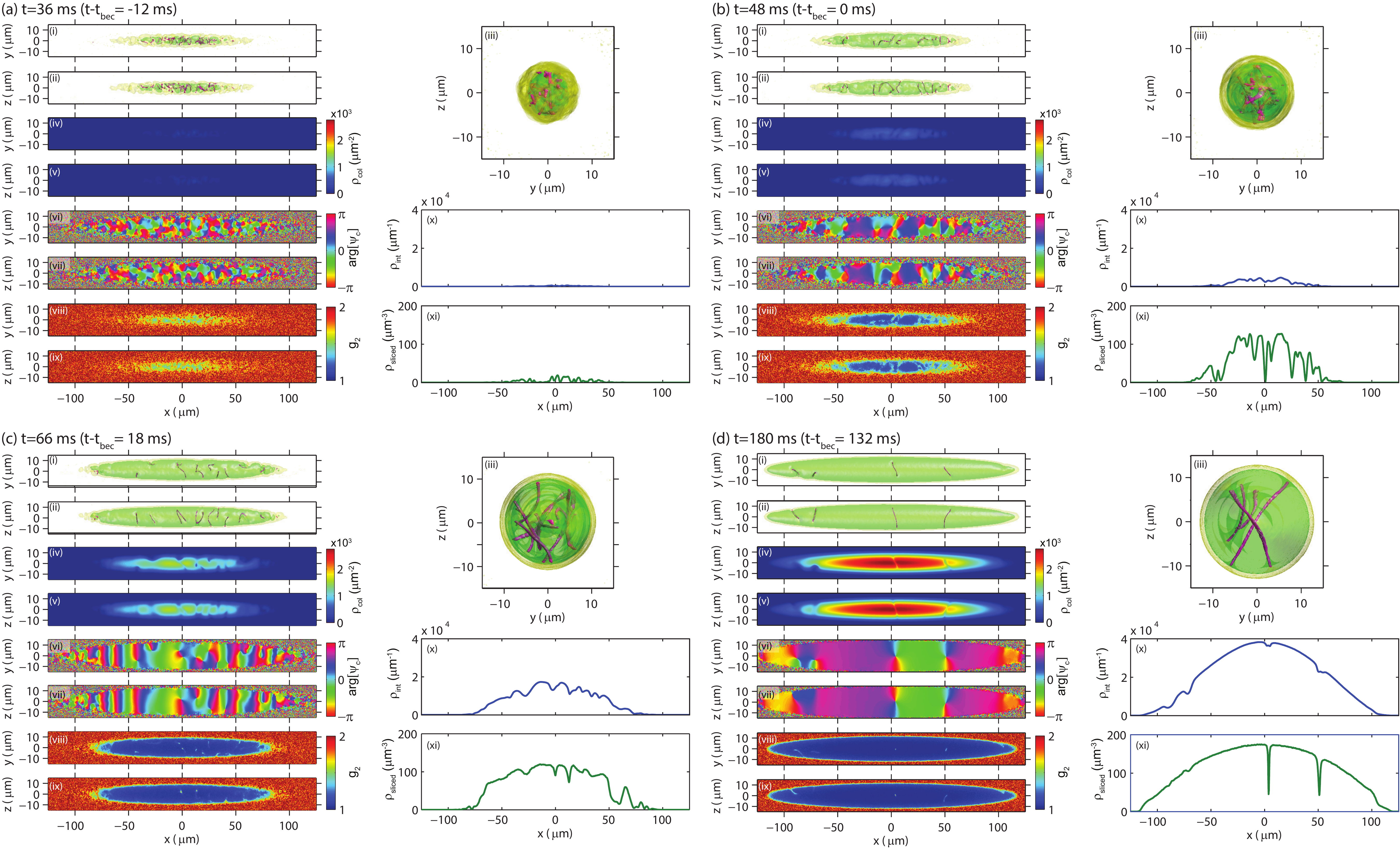}
\centering
\caption{
{\bf Crossover from Random Defects to Solitonic Vortices:}
Shown are different visualizations of density and phase of the quenched ($\tau_R =84$ ms)
mode corresponding to the largest eigenvalue at four characteristic evolution times 36, 48, 60, and 180 ms, corresponding to $t-t_{\rm bec} = -12$, $0$, $18$ and $132$ ms. 
In each panel, the top subpanels illustrate the highest-mode (Penrose-Onsager) density isosurfaces from different viewing angles set by (i) $z=0$ and (ii) $y=0$, with axial view images shown in the square panels in (iii). 
Subsequent panels (iv) and (v) depict the column-integrated densities, respectively integrated over the $z$ and $y$ axes, and the phase profiles in the (vi) $z=0$ and (vii) $y=0$ planes. 
The two bottom-left panels show the corresponding local second order correlation function, 
$g_2( \mathbf{r},\mathbf{r},\mathbf{r},\mathbf{r})$, in the (viii) $z=0$ and (ix) $y=0$ planes.
Transversally-integrated density of the Penrose-Onsager mode $\rho_{\rm int}(x) = \iint dydz n_{\rm PO}(\mathbf{r})$
 is shown in (x), with the sliced density $\rho_{\rm sliced}(x) = n_{\rm PO}(x,y=0,z=0,t)$ shown in (xi), where $n_{\rm PO}$ the Penrose-Onsager density.
This figure demonstrates clearly the initial emergence of random-type defects (which are most clearly pronounced in the density cuts), the transition of the phase profile from random to solitonic-vortex-like, and the survival of only a handful of well-defined solitonic vortices at late evolution times, propagating on top of an otherwise largely relaxed Thomas-Fermi-like condensate profile.
These images also highlight the difficulty in counting vortices at early times from any chosen integration angle. 
We note that $t_{\rm bec}(\tau_R=84 \, {\rm ms}) = 48_{-3}^{+3}$ ms.
(A more coarse-grained dynamical evolution of projected densities is shown in \href{https://youtu.be/P4xGyr4dwKI}{Supplementary Movie 4}.)}
\label{fig_5}
\end{figure*}